\documentclass{article}

\usepackage{PRIMEarxiv}

\usepackage[utf8]{inputenc} 
\usepackage[T1]{fontenc}    
\usepackage{hyperref}       
\usepackage{url}            
\usepackage{booktabs}       
\usepackage{amsfonts}       
\usepackage{nicefrac}       
\usepackage{microtype}      
\usepackage{lipsum}
\usepackage{fancyhdr}       
\usepackage{graphicx}       
\graphicspath{{media/}}     

\usepackage[normalem]{ulem}
\usepackage[arrowdel]{physics}
\usepackage{color,soul}
\usepackage{amsmath,amssymb,amsfonts}
\usepackage{caption}
\usepackage{subcaption}
\usepackage{tikz}
\usetikzlibrary{arrows}
\usetikzlibrary{plotmarks}
\usetikzlibrary{calc}
\usetikzlibrary{decorations.shapes}
\usetikzlibrary{arrows,decorations.pathreplacing}
\usetikzlibrary{fit}
\usetikzlibrary{backgrounds}
\usetikzlibrary{automata,arrows,positioning,calc}

\usepackage[most]{tcolorbox}
\definecolor{block-gray}{gray}{0.85}
\newtcolorbox{myquote}{colback=block-gray,grow to right by=-10mm,grow to left by=-10mm,
boxrule=0pt,boxsep=0pt,breakable}

\DeclareSymbolFontAlphabet{\mathrm}    {operators}
\DeclareSymbolFontAlphabet{\mathnormal}{letters}
\DeclareSymbolFontAlphabet{\mathcal}   {symbols}
\DeclareMathAlphabet      {\mathbf}{OT1}{cmr}{bx}{n}
\DeclareMathAlphabet      {\mathsf}{OT1}{cmss}{m}{n}
\DeclareMathAlphabet      {\mathit}{OT1}{cmr}{m}{it}
\DeclareMathAlphabet      {\mathtt}{OT1}{cmtt}{m}{n}

\DeclareSymbolFont{operators}   {OT1}{cmr} {m}{n}
\DeclareSymbolFont{letters}     {OML}{cmm} {m}{it}
\DeclareSymbolFont{symbols}     {OMS}{cmsy}{m}{n}

\title{From multiscale biophysics to digital twins of tissues and organs: future opportunities for \textit{in silico}  pharmacology
\thanks{\textit{\underline{Citation}}: 
\textbf{Authors. Title. Pages.... DOI:000000/11111.}} 
}

\author{
  Michael Taynnan Barros* \\
  School of Computer Science and Electronic Engineering, University of Essex\\ Colchester, UK \\
  \texttt{m.barros@essex.ac.uk} \\
   \And
  Michelangelo Paci \\
  CBIG, BioMediTech, Faculty of Medicine and Health Technology, Tampere University\\ Tampere, Finland
   \AND
  Aapo Tervonen \\
  Department of Biological and Environmental Science, Faculty of Mathematics and Science,  University of Jyväskylä\\ Jyväskylä, Finland
 \And
  Elisa Passini \\
  National Centre for the Replacement, Refinement and Reduction of Animals in Research and\\ Department of Computer Science, Oxford University\\ Oxford, UK.
   \And
  Jussi Koivumäki \\
  CBIG, BioMediTech, Faculty of Medicine and Health Technology, Tampere University\\ Tampere, Finland
    \And
  Jari Hyttinen \\
  CBIG, BioMediTech, Faculty of Medicine and Health Technology, Tampere University\\ Tampere, Finland
     \And
  Kerstin Lenk* \\
  Institute of Neural Engineering, Graz University of Technology and\\ BioTechMed\\ Graz, Austria\\
   \texttt{elenk.kerstin@gmail.com} \\
}

\begin{document}
\maketitle

\begin{abstract}
With many advancements in \textit{in silico} biology in recent years, the paramount challenge is to translate the accumulated knowledge into exciting industry partnerships and clinical applications. Achieving models that characterize the link of molecular interactions to the activity and structure of a whole organ are termed \textit{multiscale biophysics}. Historically, the pharmaceutical industry has worked well with \textit{in silico} models by leveraging their prediction capabilities for drug testing. However, the needed higher fidelity and higher resolution of models for efficient prediction of pharmacological phenomenon dictates that \textit{in silico} approaches must account for the verifiable multiscale biophysical phenomena, as a spatial and temporal dimension variation for different processes and models. The collection of different multiscale models for different tissues and organs can compose digital twin solutions towards becoming a service for researchers, clinicians, and drug developers. Our paper has  two main goals: 1) To clarify to what extent detailed single- and multiscale modeling has been accomplished thus far, we provide a review on this topic focusing on the biophysics of epithelial, cardiac, and brain tissues; 2) To discuss the present and future role of multiscale biophysics in \textit{in silico}  pharmacology as a digital twin solution by defining a roadmap from simple biophysical models to powerful prediction tools. Digital twins have the potential to pave the way for extensive clinical and pharmaceutical usage of multiscale models and our paper shows the basic fundamentals and opportunities towards their accurate development enabling the quantum leaps of future precise and personalized medical software.
\end{abstract}


\section{Introduction} 

Developing digital representations of the human body that characterizes both spatial structures as well as activity needs to account for the link between molecules, their propagation dynamics, and their reactions in  various scales, thus bringing together a set of models that are termed \textit{multiscale biophysics}.
Modern biophysics brings together physical molecular behavior with a highly complex biological activity and structure separated in multiple spatial and temporal scales, herein referred to as multiscale biophysics. Systems biology does not have a general law for the multiscale phenomena that govern most eukaryotic organisms \cite{mcculloch2016systems}. Capturing all the phenomena present in cells and tissues into a mathematical model is an arduous task due to the plurality of molecules and their respective pathways \cite{pesenson2013introduction}. Based on the number of time and space-dependent variables, most realistic approaches progress towards non-tractable non-linear dynamic models. The existing solutions for modeling multiscale biological systems rely on simulators that are data-intensive, which means that experimental data is used to tailor pre-existing simulators to fit the specific time and spatial behaviors of a particular system \cite{barros2018multi}. Even though considerably accurate, multiscale modeling is not translatable, and new models will have to go through the same time-consuming task of model creation and validation. Today, researchers of computational biology and branches thereof do not have a clear basic guideline to tackle multiscale complexity. This hinders the validation and creation of new models that would bring forth forward more holistic computational frameworks to the market. In addition, the large biological variability observed in the population brings another layer of complexity. In recent years, the importance of developing personalized computer models to enable precision pharmacology and medicine has become more and more evident.
The ambition is to make an advanced software that can be useful not only in biomedical research but also in clinical settings and pharmaceutical industries, and every other natural setting where prediction is most welcomed.
This situation can be radically changed with high-fidelity digital twins. They are defined as full digital reconstructions of tissue and organs from our bodies that exhibit detailed biological function and structure descriptions that facilitate multidimensional predictions. Digital twins will build the bridge of personalized patient-specific computational models with clinical recommendation guidelines for disease treatment and prevention by using patient data to adjust the models with lower variability and increased prediction power \cite{corral-acero_digital_2020,bjornsson2020digital,sun2022digital}. 

Multiscale models pick up the existing successful efforts of \textit{in silico}  biology \cite{walpole2013multiscale} which has been confined to specific non-scalable analysis, and that already has been proven to have a place in big pharmaceutical players \cite{passini2021virtual}. The pharmaceutical industry can leverage efforts on multiscale modeling into the realm of cell-free drug discovery and screening technology. From the human body microscales, drugs interact with specific molecules in the intracellular space and the cellular membrane that influence tissues and organs. 
Future prediction techniques on drug effects will be data-intensive and based on digital multiscale twins, where a large number of drug types will be evaluated under an integrated digital twin solution. Either a top-down or a bottom-up approach works here, where drugs shall be analyzed from molecules to organs and vice-versa. This can help to deliver safer and more effective drugs to the market in less time with a large economic impact. By identifying adverse drug effects and sub-populations at a higher risk very early on in the drug development pipeline, \textit{in silico} trials could be used to integrate or even replace the current methodologies, also contributing to an overall reduction of animal use and costs.

Although in this review, we concentrate our analysis on multiscale models tailored for the pharmaceutical industry, our characterization of multiscale biophysics can transcend this scenario and reach way beyond - across academia, industry, clinical, and regulatory settings. As an example, they can assist clinicians by providing new insights into disease mechanisms and predicting the most effective pharmacological (or non-pharmacological) intervention for a specific patient. For the heart, numerous such examples can be found in the literature \cite{passini2016mechanisms,sung2021analyzing,wang2021human,ni2020populations}. In academia, multiscale models as a multidisciplinary research direction can be very useful for individuals with little expertise in computer modeling and simulation. They also can suggest new directions for regulatory agencies like the European Medicines Agency (EMA) and the U.S. Food and Drug Administration (FDA). These examples contribute to explaining why the economic impact of \textit{in silico}  models is estimated to be USD 13.6\$ billion by 2026 \cite{WinNT}.

\begin{figure*}[!]
    \centering
    \includegraphics[width=1.0\textwidth]{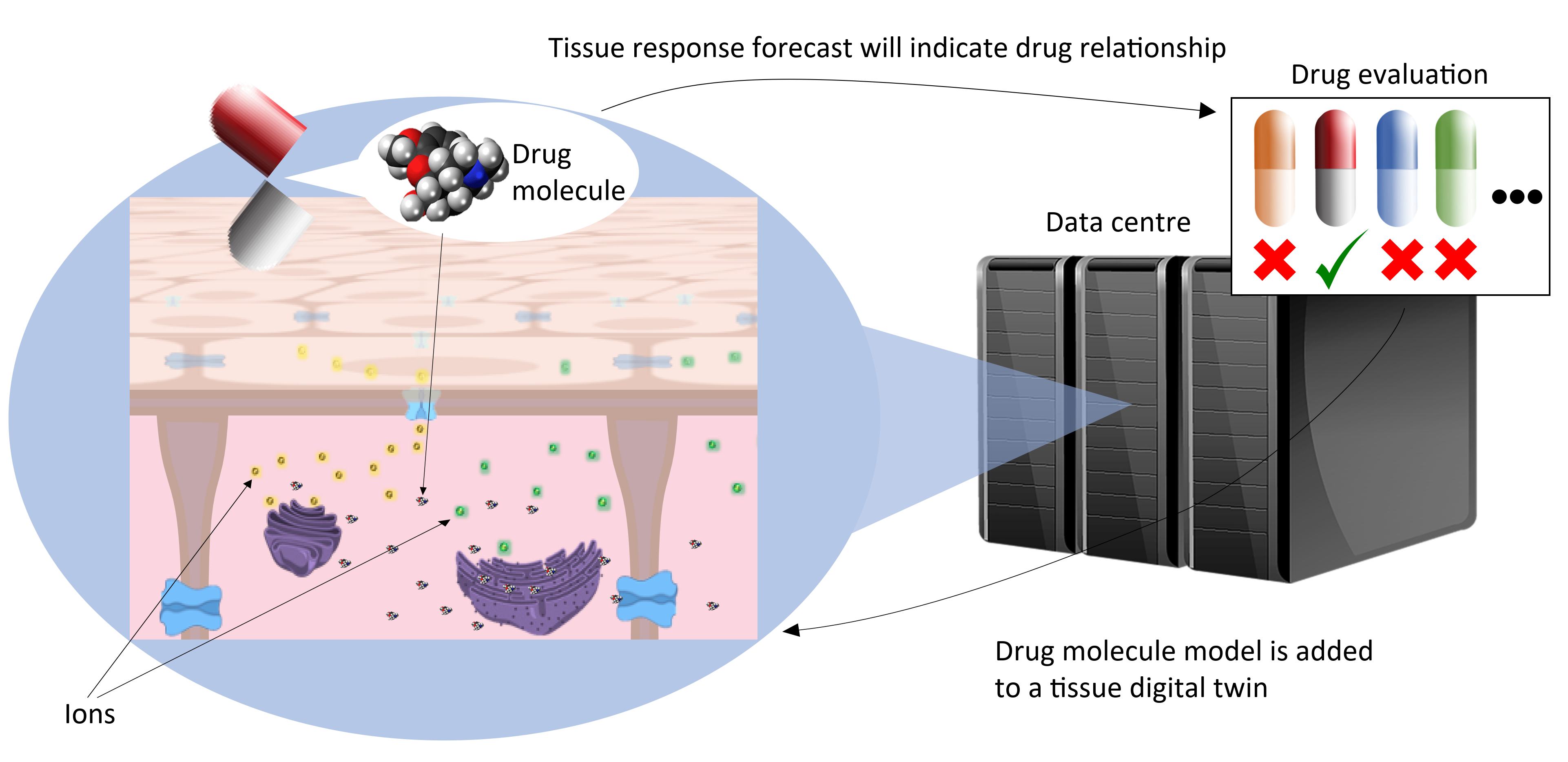}
    \caption{Digital twin tissues reconstructed through multiscale modeling will be used in data centers to perform drug safety and efficacy evaluations. The forecast response will predict potential drug-induced effects and inform subsequent experiments.}
    \label{fig:scenario}
\end{figure*}

Multiscale modeling in itself poses many challenges, as highlighted by the biophysical modeling community in many different works including \cite{fletcher2022seven,karabasov2014multiscale,hoekstra2014multiscale}. The main challenges can be summarised as follows: a) lack of good quality experimental data able to capture the dynamical behaviors at multiple scales; b) scarcity of mathematical tools that can bridge multiples scales and can be run without requiring high-performance computing; c) incomplete biophysical knowledge of the biological systems at all scales. Therefore, there is a need for a community effort to develop verified and easy-to-use simulation frameworks that will facilitate multiscale simulations for predictions of physiological activity.

In this paper, we aim to review existing works of biophysical models from single ion channels to whole organs, also including single cells, multiple cells, and tissue, across different spatial and temporal scales. The main goal is to provide a comprehensive modeling reference for either academic or industry researchers that want to learn the basics of multiscale biophysics tailored for pharmacological studies at the start or advanced stages of their careers. We explore the mathematical models in a descriptive fashion while maintaining an accessible level with relevant modeling complexity. Of course, the topic of biophysics is much broader than what is shown in this paper, however, it motivates and educates how multiscale is important and how it can be simulated. We decided to target the cardiac, brain, and epithelial tissues since we believe they provide a good example of the complexity and variability found in human biological systems. They also provide a good representation of different development statuses of \textit{in silico} modeling - ranging from cardiac models as the most advanced models in this context to brain models to epithelial models. We will show the recent advances of multiscale modeling and simulations in these fields, and the existing integrative approaches to link biological data and model development, together with their challenges, e.g., consideration of variability, reaction-diffusion concurrency, the integrity of the interaction between the scales, etc. We meet our goal by exploring computer models in each scale and drafting links between them, using an accessible mathematical description that enables an open-wide understanding of the dynamics of each system. Further, we discuss the current challenges to raise the profile of \textit{in silico}  biology as a reliable source for prediction. Finally, we highlight the impact of multiscale modeling and simulations in pharmacology to show how they can reshape the landscape of existing solutions and provide support across academia, industry, clinicians, and regulators.








\section{Multi-scale biophysics in the body for digital twins solutions} 


We base our analysis on four to six bottom-up spatial and temporal scales depending on the considered tissue type: the sub-cellular, cellular, cell-cell, tissue, tissue-tissue, and the whole organ level (Figs. \ref{fig:rule_heart}, \ref{fig:rule_brain}, and \ref{fig:rule_ep}). We do not include proteomics or genomics in this paper due to their additional levels of complexity. We concentrate on modeling approaches varying from typical biophysical approaches (including ordinary differential equations, partial differential equations, reaction-diffusion, and continuum methods) and phenomenological models (stochastic, rate-based, and cellular mass models) and show how existing literature deals with the rising complexity and volume of data when considering rich biophysics with molecular dynamics in modeling. 
We have selected the following models based on the analyses of the literature explored in Appendix A. Now, in the following, we will dive into the biophysical models from each scale for the cardiac, brain, and epithelial tissues. 




\subsection{Cardiac models}

\begin{figure}
    \centering
    \includegraphics[width=\textwidth]{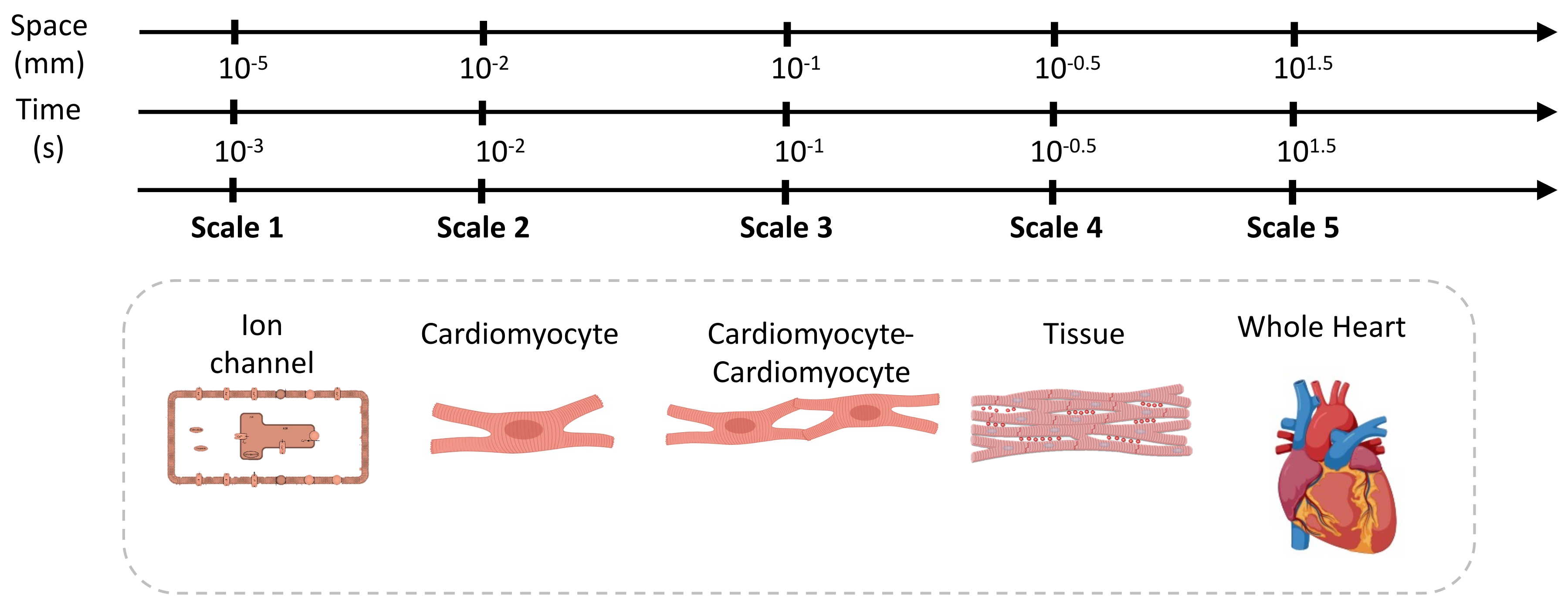}
    \caption{Five different spatial and temporal scales that are linked with the progression from intracellular pathways to organs in cardiac models.}
    \label{fig:rule_heart}
\end{figure}
 
The heart, and in particular cardiac electrophysiology, is currently the domain where multiscale modeling demonstrates close to its full potential. The existing cardiac models constitute a reference example for other systems of the body. The conceptual pipeline is, in principle, straightforward, and models at the higher scales rely upon the lower levels (Fig. \ref{fig:rule_heart}): 1) single ion channels/currents/transports, 2) single cardiomyocytes (CM), 3) tissue, and 4) whole organ \cite{pathmanathan2018validation}. Furthermore, each scale can contain additional levels of complexity, e.g., modeling single ionic currents using detailed paradigms (Hodgkin-Huxley, Markov \cite{rudy2006computational}, molecular-structure-to-function \cite{ramasubramanian2018structural}) or simulating different CM phenotypes (ventricular \cite{bartolucci2020simulation} vs. atrial \cite{koivumaki2014silico}). 
Advanced digital reconstructions of the heart, including the Living Heart Project \cite{baillargeon2014living}, are perfect examples of how the multiscale can be understood. However, there are still existing challenges in cardiac modeling that have only been  partially addressed, e.g., patient-profile-based tissue variability, and disease modeling, as described in the work of McCulloch \cite{mcculloch2016systems}. 
In this section, we will clarify what modeling work is needed to perform simulations from single-cell to the whole organ level, taking as an example cardiac electrophysiology. We present the conceptual pipeline, also including some mathematical details. However, given the extension and complexity of the topic, we refer to previous works for an advanced description \cite{hille2001electrophysiology, rudy2006computational, pullan2005multiscale}.
The trustworthiness and validation of cardiac models are  important points and have been previously analyzed \cite{pathmanathan2018validation}. They will not be further discussed in this paper.

\subsubsection{Single ion currents - The Hodgkin-Huxley paradigm (scale 1)}
\begin{figure}
    \centering
    \includegraphics[width=0.45\textwidth]{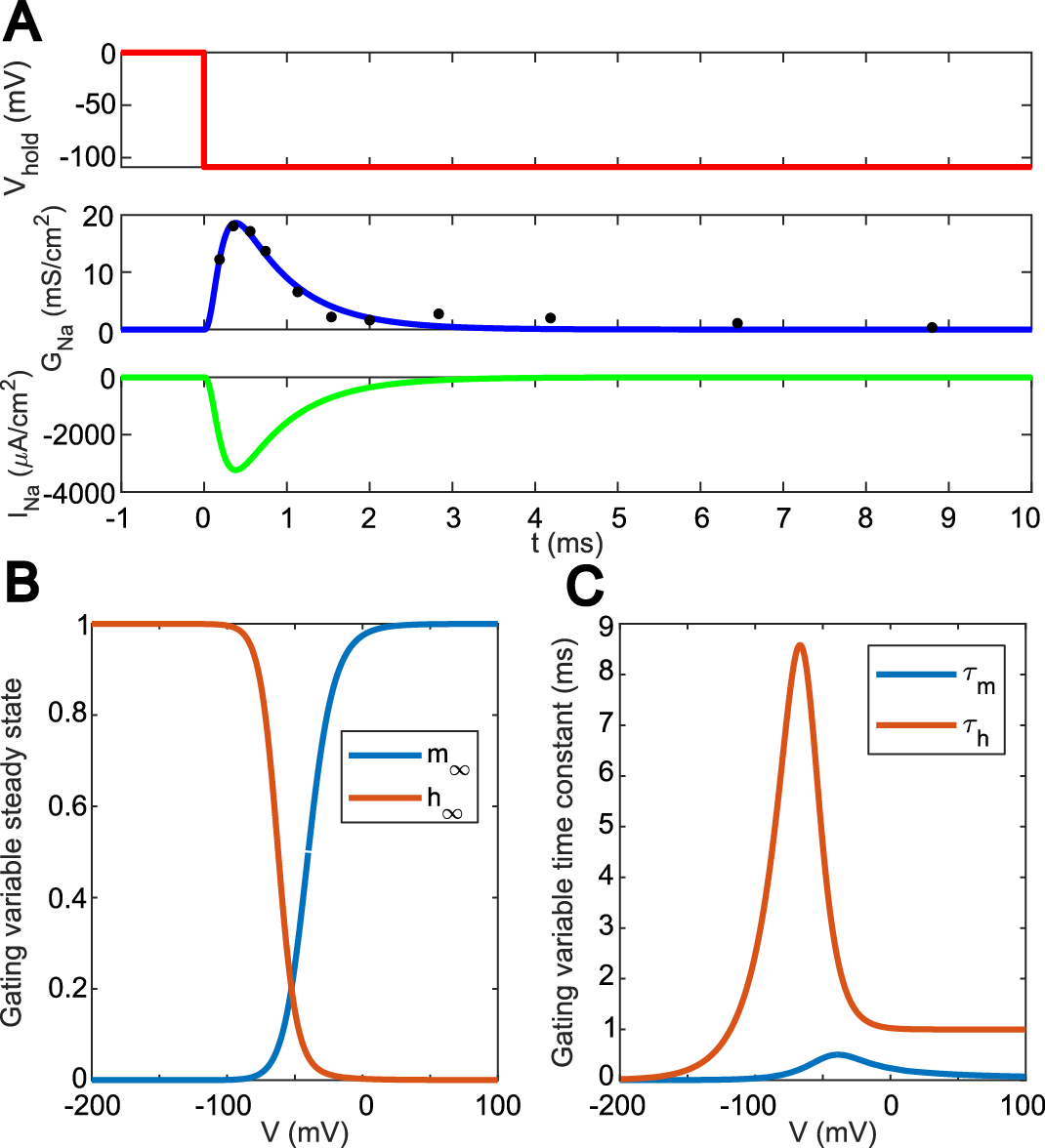}
    \caption{Simulation of cardiac ionic currents. (A) Simulation of a voltage-clamp step with the Hodgkin-Huxley model. From top to bottom: holding potential -109 mV (red); timecourse of $I_{Na}$ conductance (blue) with \textit{in vitro} experiments (black dots) from \cite{hodgkin1952simulation}; timecourse of $I_{Na}$ (green). (B) $I_{Na}$ steady state activation ($m_{\infty}$) and inactivation ($h_{\infty}$) curves. (C) $I_{Na}$ time constants of activation ($\tau_{m}$) and inactivation ($\tau_{h}$).}
    \label{fig:fig_pax_HH}
\end{figure}

The most basic brick we consider in this section is the single ion current flowing through the cell membrane via selective ion channels, given the difference of potential $V_m$ at the two membrane sides.
\begin{equation} \label{eq:resistive_current}
I_{ion} = G_{ion}(V_m - E_{ion}),
\end{equation}
where $G_{ion}$ represents the constant conductance of the current, usually expressed in $nS/pF$, and $E_{ion}$ is the Nernst equilibrium potential for the $ion$ species computed as
\begin{equation} \label{eq:nernst_potential}
E_{ion} = \frac{RT}{zF}ln(\frac{[ion]_o}{[ion]_i}),
\end{equation}
where R is the universal gas constant, T is the temperature, F is Faraday’s constant, and $[ion]_i$ and $[ion]_o$ are the intracellular and extracellular concentrations.
Such a simple formalism is not adequate for simulating more complex current evolution in time. For example, when $I_{Na}$ is triggered by a voltage step, it shows first a rapid activation and then a slow inactivation (Fig. \ref{fig:fig_pax_HH} A). Mathematical constructs like the activation and inactivation gating variables enable the simulation of a voltage- and time-dependent modulation of the current conductance. In case of $I_{Na}$, the equation \ref{eq:resistive_current} changes into 
\begin{equation} \label{eq:INaHH}
I_{Na} = G_{Na}(V_m - E_{Na}) = \overline{G_{Na}}m^3h(V_m - E_{Na}).
\end{equation}
In this equation, $G_{Na}$ conductance is composed of three terms: the constant maximum conductance $\overline{G_{Na}}$, the series of three activation gating variables $m$ and the inactivation gating variable $h$. A generic gating variable $x(t, V_m)$ can be represented in two equivalent forms. The first formulation
\begin{equation} \label{eq:gating_x}
\frac{dx}{dt} = \alpha_x (1-x)-\beta_x x
\end{equation}
highlights the meaning of $x$ and $1-x$ as the open and closed probabilities of the gate $x$. If the values of either $\alpha_x$ or $\beta_x$ depend on $V_m$, $x$ is both voltage-dependent and time-dependent.
By imposing
\begin{equation} \label{eq:gating_alpha}
\alpha_x = \frac{x_\infty}{\tau_x}
\end{equation}
and 
\begin{equation}\label{eq:gating_beta}
\beta_x = \frac{1-x_\infty}{\tau_x}
\end{equation}
equation \ref{eq:gating_x} can be written as
\begin{equation} \label{gating_x_2}
\frac{dx}{dt} = \frac{x_\infty-x}{\tau_x}
\end{equation}
 in order to highlight the dynamic sense of the gating variable $x(t, V_m)$.
The steady-state
\begin{equation} \label{eq:steady-state}
x_\infty = \frac{\alpha_x}{\alpha_x + \beta_x}
\end{equation}
represents the value to which $x$ tends at a certain voltage and it is  obtained by fitting the \textit{in vitro} data using the sigmoid function
\begin{equation} \label{eq:gating_sigmoid}
x_\infty = \frac{1}{1+e^\frac{V_m - V_{h}}{K}},
\end{equation}
where $V_{h}$ is the voltage of half activation and $K$ is the gradient of activation.  
The time constant
\begin{equation} \label{eq:gating_tau}
\tau_x = \frac{1}{\alpha_x + \beta_x}
\end{equation}
represents how fast the gating variable $x$ reaches its steady-state $x_\infty$ (Fig. \ref{fig:fig_pax_HH} B and C).
Following this modeling paradigm, the conductance $G_{Na}(t, V_m)$ is a function of the open probabilities of the series of three voltage- and time-dependent activation and one inactivation gating variables. Although very powerful, this representation of an ion current kinetics suffers two important limitations. First, the gates are not related to any actual kinetic state of the ion channels. Second,  the activation and inactivation gates are independent. For example, regarding $I_{Na}$, it has been shown that its inactivation is more likely to happen when the channel is open \cite{armstrong1977electrophysiology, bezanilla1977electrophysiology}. This cannot be captured by modeling $I_{Na}$ with the Hodgkin-Huxley paradigm.

\subsubsection{Single ion currents - The Markov paradigm (scale 1)}
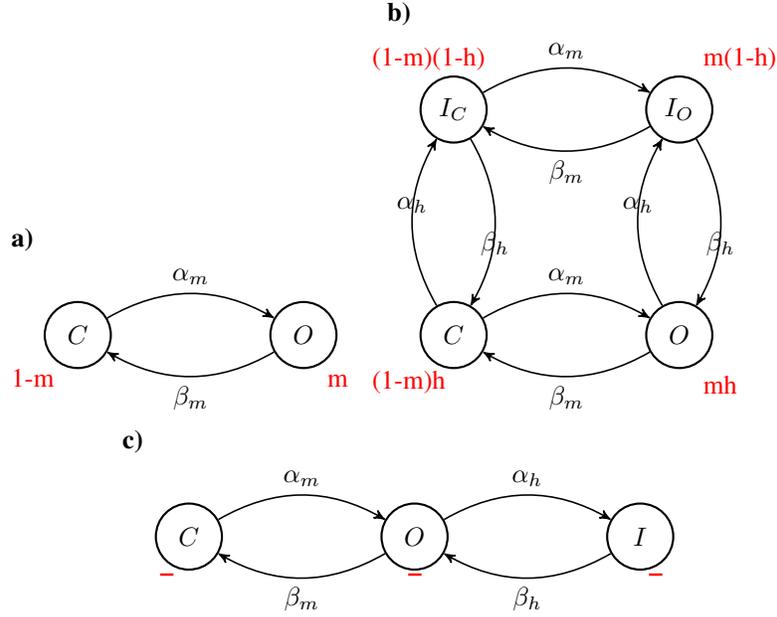
\begin{figure}
    \centering
    \begin{tikzpicture}[->, >=stealth', auto, semithick, node distance=3cm]
	\tikzstyle{every state}=[fill=white,draw=black,thick,text=black,scale=1]
	\node[state]    (A)                     {$C$};
	\node[state]    (B)[right of=A]   {$O$};
	\path
	(A) edge[bend left,above]			node{$\alpha_m$}	(B)
	(B) edge[bend left,below]	node{$\beta_m$}	(A);
	\path[fill=red,line width=0.056pt] (-1,-0.8) node[above right]
    (text1190) {\color{red} 1-m};
    \path[fill=red,line width=0.056pt] (3.2,-0.8) node[above right]
    (text1190) {\color{red} m};
    \path[fill=red,line width=0.056pt] (-1,1) node[above right]
    (text1190) {\bf a)};
	\end{tikzpicture}
	\begin{tikzpicture}[->, >=stealth', auto, semithick, node distance=3cm]
	\tikzstyle{every state}=[fill=white,draw=black,thick,text=black,scale=1]
	\node[state]    (A)                     {$I_C$};
	\node[state]    (B)[right of=A]   {$I_O$};
	\node[state]    (C)[below of=A]   {$C$};
	\node[state]    (D)[below of=B]   {$O$};
	\path
	(A) edge[bend left,above]			node{$\alpha_m$}	(B)
	(B) edge[bend left,below]	node{$\beta_m$}	(A)
	(C) edge[bend left,above]			node{$\alpha_h$}	(A)
	(A) edge[bend left,below]	node{$\beta_h$}	(C)
	(D) edge[bend left,above]			node{$\alpha_h$}	(B)
	(B) edge[bend left,below]	node{$\beta_h$}	(D)
	(C) edge[bend left,above]			node{$\alpha_m$}	(D)
	(D) edge[bend left,below]	node{$\beta_m$}	(C);
	\path[fill=red,line width=0.056pt] (-1,1) node[above right]
    (text1190) {\bf b)};
    \path[fill=red,line width=0.056pt] (-1.2,0.4) node[above right]
    (text1190) {\color{red} (1-m)(1-h)};
    \path[fill=red,line width=0.056pt] (3.2,0.4) node[above right]
    (text1190) {\color{red} m(1-h)};
    \path[fill=red,line width=0.056pt] (-1.2,-3.9) node[above right]
    (text1190) {\color{red} (1-m)h};
    \path[fill=red,line width=0.056pt] (3.2,-3.9) node[above right]
    (text1190) {\color{red} mh};
	\end{tikzpicture}\\
	\begin{tikzpicture}[->, >=stealth', auto, semithick, node distance=3cm]
	\tikzstyle{every state}=[fill=white,draw=black,thick,text=black,scale=1]
	\node[state]    (A)                     {$C$};
	\node[state]    (B)[right of=A]   {$O$};
	\node[state]    (C)[right of=B]   {$I$};
	\path
	(A) edge[bend left,above]			node{$\alpha_m$}	(B)
	(B) edge[bend left,below]	node{$\beta_m$}	(A)
	(B) edge[bend left,above]			node{$\alpha_h$}	(C)
	(C) edge[bend left,below]	node{$\beta_h$}	(B);
	\path[fill=red,line width=0.056pt] (-1,1) node[above right]
    (text1190) {\bf c)};
    \path[fill=red,line width=0.056pt] (-0.5,-0.7) node[above right]
    (text1190) {\color{red} \bf --};
    \path[fill=red,line width=0.056pt] (2.8,-0.7) node[above right]
    (text1190) {\color{red} \bf --};
    \path[fill=red,line width=0.056pt] (6,-0.7) node[above right]
    (text1190) {\color{red} \bf --};
	\end{tikzpicture}
    \caption{Three illustrative examples of Markov models. In red, we report the occupancy probability for the Markov states. For the open states $O$ in panels A and B, the occupancy probability for the Markov states corresponds to the open probability of an equivalent Hodgkin-Huxley (HH) model. (A) Two-state model with an open $O$ and a close $C$ state. This model corresponds to a single HH activation gate. (B) Four-state model with one open $O$, one close $C$, and two inactivated $I_O$ and $I_C$ states. Since the transition rates  $\alpha_m$ and $\beta_m$ are the same in the horizontal transitions, they can be represented with a single HH gating variable $m$. The same applies to $\alpha_h$ and $\beta_h$. In this model, the activation is independent of the activation, so an equivalent HH model with two gating variables $m$ and $h$ exists. (C) Three-state model with one open $O$, one close $C$ and one inactivated $I$ state. Here, inactivation is dependent on activation, therefore no HH equivalent model exists.}
    \label{fig:fig_pax_Markov}
\end{figure}

A more powerful modeling paradigm, that can represent the dependence of a given transition on the occupancy of different biophysical states of the channel, is the Markov paradigm. Markov models take into consideration multiple channel conformation states and state-to-state transitions, which have been characterized \textit{in vitro}. This type of model considers that one transition between channel states depends on the present conformation of the channel, but not on previous behavior \cite{rudy2006computational}.
As the first example, we take a look at a channel characterized only by a single open (O) and a single closed (C) state (Fig. \ref{fig:fig_pax_Markov}A).
The rate of occupancy of the two states is expressed by the following first-order equations
\begin{eqnarray} 
    \label{eq:markov-2-states-C}
    \frac{dC}{dt} & = & -\alpha_mC+\beta_mO  \\
    \label{eq:markov-2-states-O}
    \frac{dO}{dt} & = &  \alpha_mC-\beta_mO
\end{eqnarray}
Considering that the open Markov state corresponds to the probability that the channel is in the open state  $O = P_{open} = m$ and similarly for the closed state $C = P_{closed} = 1-m$, Eq. \ref{eq:markov-2-states-O} can be written as
\begin{equation} \label{eq:gating_m}
\frac{dm}{dt} = \alpha_m (1-m)-\beta_m m
\end{equation}
that is nothing more than the formulation of the activation gating variable $m$ in the Hodgkin-Huxley $I_{Na}$ according to Eq. \ref{eq:gating_x}. However, ion channels can also be inactivated. We consider a slightly more complex Markov model with four states  closed ($C$), open ($O$), open inactivated ($I_O$), and closed inactivated ($I_C$), and transitions rates equal in the horizontal sides and in the vertical sides of the square diagram in Fig. \ref{fig:fig_pax_Markov}A. This model can be expressed by the following first-order equations
\begin{eqnarray} 
    \label{eq:markov-4-states-C}
    \frac{dC}{dt} & = & \beta_m O + \beta_h I_C - (\alpha_m + \alpha_h) C\\
    \label{eq:markov-4-states-O}
    \frac{dO}{dt} & = &  \alpha_m C + \beta_h I_O - (\beta_m + \alpha_h) O\\
    \label{eq:markov-4-states-IO}
    \frac{dI_O}{dt} & = & \alpha_h O + \alpha_m I_C -(\beta_h + \beta_m) I_O\\
    \label{eq:markov-4-states-IC}
    \frac{dI_C}{dt} & = &  \alpha_h C + \beta_m I_O -(\alpha_m + \beta_h) I_C\\
\end{eqnarray}
The fact that the horizontal transition rates $\alpha_m$ (rate of activation) and $\beta_m$ (rate of deactivation) are the same for $C-O$ and $I_C-I_O$ means that they can be represented by a single gate $m$ as in the previous example. Applying the same approach for the vertical transition rates $\alpha_h$ (rate of inactivation) and $\beta_h$ (rate of recovery) and sides, we can obtain the following equation similar to Eq. \ref{eq:gating_m}
\begin{equation} 
\frac{dh}{dt} = \alpha_h (1-h)-\beta_h h
\end{equation}
that again is nothing more than the formulation of the inactivation gating variable $h$ in the Hodgkin-Huxley $I_{Na}$. Furthermore, since the inactivation rates are the same on the vertical side, the inactivation of the channel is independent of the states $C$ and $O$ or, in other terms, is independent of the activation gate $m$, as in the Hodgkin-Huxley $I_{Na}$. Considering again the state occupancy probabilities, we can write $C = (1-m)h$, $I_C = (1-m)(1-h)$, $I_O = m(1-h)$ and $O = mh$. In particular, the probability that the channel is in the $O$ state closely resembles the gating variable product in Eq. \ref{eq:INaHH} (except for the third power). The previous considerations demonstrate that it is possible to convert specific Markov models with certain geometries and independent activation and inactivation into Hodgkin-Huxley models.
Since an ionic current is a sum of currents flowing through multiple channels, an equivalent interpretation of a Markov state is that it is the ratio between the channels in that specific state and the total number of channels. Markov models are very intuitive in that they obey the conservation law: the sum of the states must be 1 and each state can assume values $[0, 1]$. Therefore, a Markov model can be represented in a reduced form, for example, replacing Eq. \ref{eq:markov-4-states-IC} by 
\begin{equation}
    \label{eq:markov-reduced-IC}
    I_C = 1 - (C + O + I_O)
\end{equation}
However, not all the Markov models can be translated into Hodgkin-Huxley models. This is clear, e.g., in the case of inactivation dependent on the activation, as in the simple example proposed by \cite{rudy2006computational} and illustrated in Fig. \ref{fig:fig_pax_Markov}C. In this three-state model, the inactivated state $I$ can be reached only from the open state $O$ and the assumption of independent gating is not valid. Independently of the complexity of the Markov model for a channel, the form of the macroscopic ion current, in this case of $I_{Na}$ is the following
\begin{equation}
    \label{eq:markov-INa- simple}
    \begin{aligned}
    I_{Na} = G_{Na}(V_m - E_{Na}) = \overline{G_{Na}}O(V_m - E_{Na})\\
    = \overline{g_{sc,Na}}nO(V_m - E_{Na})
    \end{aligned}
\end{equation}
where $\overline{g_{sc,Na}}$ is the single channel conductance and $n$ is the number of channels. 


\subsubsection{Modelling a single CM (scale 2)}
\begin{figure}
    \centering
    \includegraphics[width=0.45\textwidth]{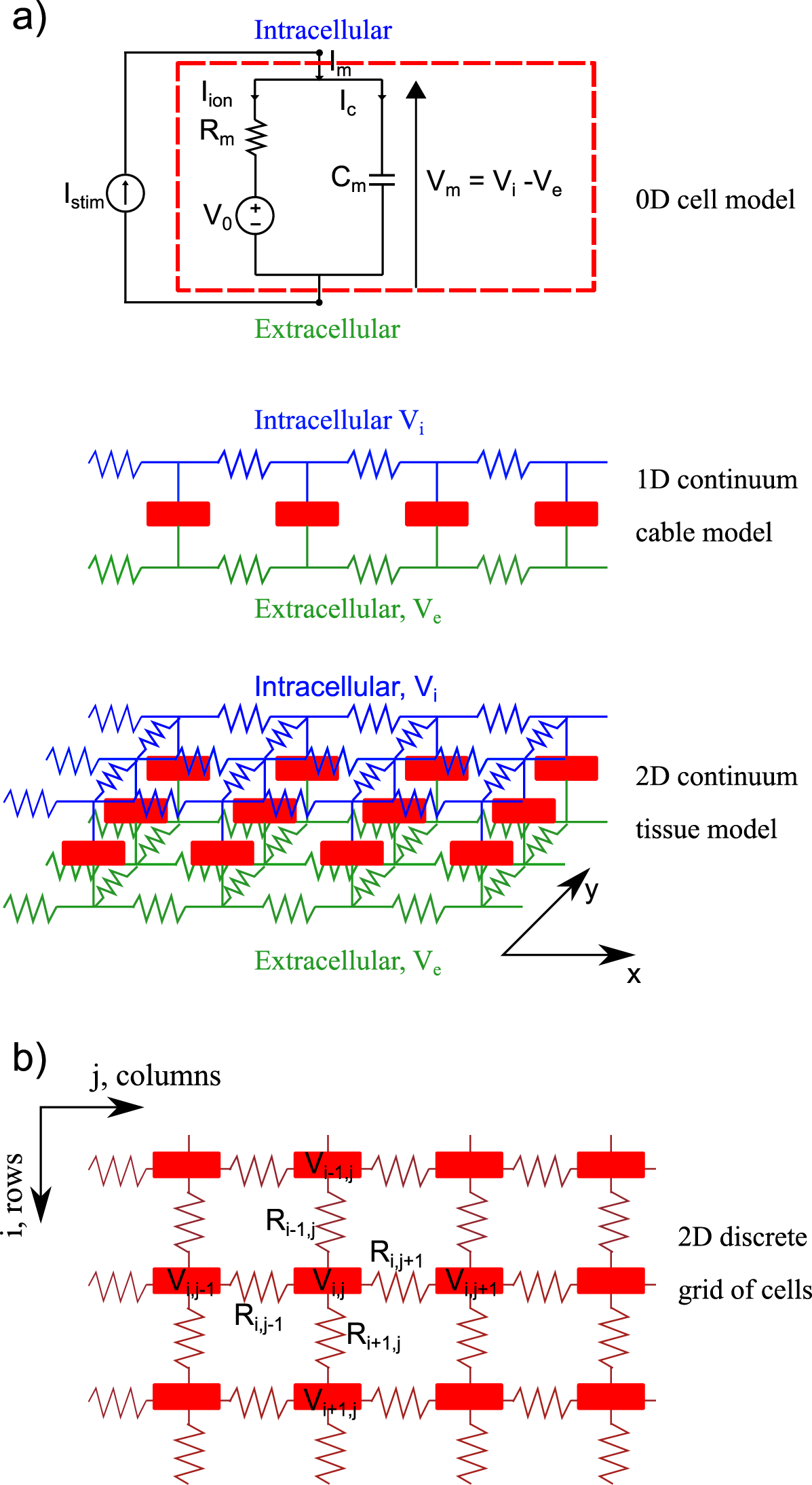}
    \caption{
    From single cell to tissue models. (A) Continuum membrane model. One cell membrane element (included in the red rectangle in the top panel) is represented as the $R_mC_m$ circuit, where $R_m$ is the resistive behavior of the cell membrane due to the ion channels that allow the passage of ions, converse to $C_m$, i.e., the capacitive behavior due to the double lipid bilayer. A sequence of membrane elements can be connected together to simulate a cable (middle panel) and a tissue (bottom panel). The intracellular (cyan) and extracellular (green) spaces are represented as a grid of resistors for electrical continuity. Continua models do not simulate individual CMs. (B) The discrete tissue model. Each single CM model (red rectangles) is explicitly electrotonically coupled with its neighbors via gap junctions (magenta resistors).}
    \label{fig:fig_pax_membr_equiv}
\end{figure}

In the previous sections, we introduced two powerful modeling paradigms, namely Hodgkin-Huxley and Markov, to simulate ion currents. As in the seminal work of Hodgkin and Huxley \cite{hodgkin1952simulation} the three ion currents $I_{Na}$, $I_K$, and $I_L$ were gathered to simulate the initiation and propagation of the action potential (AP) in the squid axon, the same approach has been used to simulate also the cardiac AP in a single CM, using the well-known equation  
\begin{equation}
    \label{eq:HH-dVdt}
\frac{\mathrm{d} V_{m}}{\mathrm{~d} t}=-\frac{I_{ion} - I_{stim}}{C_m},
\end{equation}
where $V_m = V_i - V_e$ represents the cell membrane potential computed as the difference between the intra- and extracellular potential, $I_{ion}$ the sum of the ion currents flowing through the ion channels, active transports, etc. (e.g., $I_{ion} = I_{Na}+I_K+I_L$ in the Hodgkin-Huxley model), $I_{stim}$ the stimulus/pacing current and $C_m$ the membrane capacitance.
Eq. \ref{eq:HH-dVdt} highlights the capacitive and resistive nature of the cell membrane, as shown in Fig. \ref{fig:fig_pax_membr_equiv}A, first row.







Although a detailed review of the goals, pros and cons, and impact of each CM model from the first Noble model in 1962 \cite{noble1962simulation} up to the state-of-the-art ones is out of the scope of this paper, we acknowledge that this plethora of models can be divided into two generations. The first generation includes the most seminal CM models like the aforementioned Noble (1962, three ion currents) and McAllister-Noble-Tsien (1975, nine ion currents) \cite{mcallister1975simulation} of generic Purkinje cells and the first ventricular CM Beeler-Reuter (1977, four ion currents) model \cite{beeler1977simulation}. However, it is with the second generation of models that more complex mechanisms started being included. For example, the first second-generation model by Di Francesco \textit{et al.} (1985) \cite{difrancesco1985simulation} has mechanisms that transcend Eq. \ref{eq:HH-dVdt} or the current formulations we presented in the previous sections. In fact, it includes i) a sarcoplasmic reticulum (SR) separated from the cytosol and divided into one uptake and one release store, ii) the release and uptake fluxes moving $Ca^{2+}$ in and out of SR, iii) a detailed description of the $Na^+$, $Ca^{2+}$ and $K^+$ ionic concentrations formulated (for the $i$-th compartment with volume $V_i$) as 
\begin{equation}
    \label{eq:diFrancesco-INaK}
\frac{d[ion]_i}{dt} = \frac{1}{FV_i}\sum_n I_{ion, n}
\end{equation}
that enable simulating also iv) ion exchangers like the $Na^+/K^+$ pump ($I_{NaK}$) and the $Na^+/Ca^{2+}$ exchanger ($I_{NCX}$). Although, these new elements require mathematical formulations beyond Eq. \ref{eq:HH-dVdt}, it still governs the simulation of the time course of the membrane voltage. For example, despite the following formulations (whose parameters are explained in detail in the original publication \cite{difrancesco1985simulation})
\begin{equation}
    \label{eq:diFrancesco-INaK2}
I_{NaK} = \overline{I_{NaK}}\frac{[K]_o}{K_{m,K}+[K]_o}\frac{[Na]_i}{K_{m,Na}+[Na]_i}
\end{equation}
or
\begin{equation}
    \label{eq:diFrancesco-INCX}
I_{NCX} = k_{NCX}\frac{e^{\frac{\gamma V_m F}{RT}}[Na]_i^3[Ca]_o-e^{\frac{(\gamma-1)V_mF}{RT}}[Na]_o^3[Ca]_i}{1+d_{NCX}([Ca]_i[Na]_o^3+[Ca]_o[Na]_i^3)}
\end{equation}
are surely different from the Hodgkin-Huxley or Markovian paradigm, $I_{NaK}$ and $I_{NCX}$ can be included into Eq. \ref{eq:HH-dVdt} in the term $I_{ion}$.

\subsubsection{Cardiac tissue modeling (scale 4)}

In this section, extending from the single CM models, we present two different approaches, one continuous and the other discrete, to simulate cardiac electrophysiology in a 2D portion of cardiac tissue. The electrical continuity in the  tissue is granted by gap junctions which are intercellular channels connecting the cytosol of adjacent individual CMs. They allow the passage of molecules and ions \cite{kanno2011electrophysiology}, and thus the propagation of the electrical signal in the tissue.
The bidomain model does not simulate individual CMs and their connections via gap junctions as discrete entities \cite{roth1992ccomputational}. Conversely, it averages the CM electrical behavior in a continuum model derived from the multidimensional expansion of the traditional cable model \cite{hodgkin1952simulation}. The "bidomain" description refers to the two domains represented in green and cyan in the third row of Fig.~ \ref{fig:fig_pax_membr_equiv}A, i.e., the extracellular and the intracellular domains, both modeled as grids of resistors. The two domains are coupled by a layer of the cell membrane, represented as red boxes, each one containing the resistive and capacitive branches of the 0D model. As reported by Roth \textit{et al.} \cite{roth1992ccomputational}, under the assumption that the spatial scale where the intra- and extra-cellular electrical gradients of interest develop is large enough compared to the CM size, this circuit can be modeled with continuum equations. With respect to the 1D cable, the bidomain model can take into account the anisotropy of the cardiac tissue: conduction velocity is anisotropic and its orientation is determined by multiple factors, like the CM direction, shape and size, excitability, the gap junction distribution and heterogeneity in the tissue \cite{kotadia2020simulation}. A full description of how to obtain the two bidomain equations, available at \cite{pullan2005multiscale}, is out of the scope of this paper; here we present their final form:
\begin{equation}
    \label{eq:bidomain_1}
\grad \cdot ((\sigma_i+\sigma_e)\grad V_e) = -\grad \cdot (\sigma_i \grad V_m) + I_{S1}
\end{equation}
\begin{equation}
\label{eq:bidomain_2}
\grad \cdot (\sigma_i \grad V_m) + \grad \cdot (\sigma_i \grad V_e) = \beta (C_m \frac{\partial V_m}{\partial t}+I_{ion})-I_{S2}
\end{equation}
where $\sigma_i$ and $\sigma_e$ are the conductivity tensors in the intra- and extracellular domains, $\beta$ is the surface-to-volume ratio of the CM membrane, $I_{S1}$ and $I_{S2}$ the stimulus current in either domain, and $V_m$, $V_e$, $C_m$ and $I_{ion}$ have the same meaning as in the previous Section. In particular, $I_{ion}$ is the sum of all the currents computed by a single CM model (see the previous Section).
At each numerical integration step, Eq. \ref{eq:bidomain_1} allows computing the extracellular potential $V_e$ given the transmembrane potential $V_m$, while Eq. \ref{eq:bidomain_2} returns $V_m$. The anisotropy of the cardiac tissue is described by the two conductivity tensors $\sigma_i$ and $\sigma_e$: in the case of a 2D model, the two spatial dimensions represent the directions longitudinal and transverse to the cardiac fibers. Several values for the $\sigma_i$ and $\sigma_e$ elements are available \cite{roth1992ccomputational, potse2006simulation}. However, there is consensus on the conductivity ratios along the longitudinal and transverse directions: $\sigma_{iL}/\sigma_{iT} \sim 10$ and $\sigma_{eL}/\sigma_{eT} \sim 2$, i.e., the intracellular domain is more anisotropic than the extracellular. However, assessing the values in actual living heart tissue is far from solved as it is an  ill-posed problem with several technological and mathematical challenges \cite{kamalakkannan2021improving}.

A simplification of the bidomain model is possible under the assumption that the intra- and extracellular domains are equally anisotropic, i.e., $\sigma_i = k\sigma_e$. In this case, the monodomain model is represented as
\begin{equation}
\label{eq:monodomain}
\frac{1}{(1+k)\beta}\grad \cdot (\sigma_i \grad V_m) = C_m \frac{\partial V_m}{\partial t}+I_{ion}-I_{S}
\end{equation}
Considering the domain $H$, its boundary  $\partial H$ and the position vector $\vec{r}$, the traditional boundary conditions \cite{krassowska1994computational} are
\begin{equation}
\label{eq:bidomain_bound_1}
(\sigma_i \grad V_m)\cdot \hat{n} = -(\sigma_i \grad V_e)\cdot \hat{n} \hfill \vec{r} \in \partial H
\end{equation}
and
\begin{equation}
\label{eq:bidomain_bound_2}
(\sigma_e \grad V_e)\cdot \hat{n} = 0 \hfill \vec{r} \in \partial H 
\end{equation}
if $H$ is immersed in a non-conductive medium. Conversely, if an extramyocardial domain is considered, e.g., a torso model $T$ with its tensor and potential $\sigma_t$ and $V_t$, Eq. \ref{eq:bidomain_bound_2} must take into account the current balance between $H$ and $T$, thus becoming 
\begin{equation}
\label{eq:bidomain_bound_3}
(\sigma_e \grad V_e)\cdot \hat{n_e} = -(\sigma_t \grad V_t)\cdot \hat{n_t} \hfill \vec{r} \in \partial H 
\end{equation}

A different and somehow more intuitive modeling approach represents a 2D cardiac tissue as a grid of single CMs, coupled together with resistors (as in Fig. \ref{fig:fig_pax_membr_equiv}C) that simulate the gap junctions. Given the CM with indexes $i$ and $j$, $CM_{i,j}$, its membrane potential can be computed as
\begin{equation}
\label{eq:grid_of_CMs_1}
\begin{split}
C_m\frac{dV_{i,j}}{dt} & = -I_{ion,i,j} + I_{i,j-1} -I_{i,j+1} +I_{i-1,j} -I_{i+1,j} \\
& = -I_{ion,i,j} +\frac{V_{i,j-1}-V_{i,j}}{R_{i,j-1}} -\frac{V_{i,j}-V_{i,j+1}}{R_{i,j+1}}\\ 
&\quad +\frac{V_{i-1,j}-V_{i,j}}{R_{i-1,j}} -\frac{V_{i,j}-V_{i+1,j}}{R_{i+1,j}}
\end{split},
\end{equation}
where $I_{ion,i,j}$ is the current computed by the chosen single cell model, as in Eq. \ref{eq:HH-dVdt}, and  $I_{i,j-1}$, $I_{i,j+1}$, $I_{i-1,j}$ and $I_{i+1,j}$ are the currents entering/leaving $CM_{i,j}$ via the surrounding gap junctions represented by the resistors $R_{i,j-1}$, $R_{i,j+1}$, $R_{i-1,j}$ and $R_{i+1,j}$. With respect to the bidomain model, this discrete approach enables to include the cell variability, e.g., by sampling each cell model parameter and modulating the gap junction resistance. For example, $CM_{i,j}$ and its neighbors can have slightly different values for their maximum conductances, which were previously sampled from \textit{in vitro} experimental ranges. One more difference is that this discrete approach can be solved with ordinary differential equations, instead of partial differential equations. In case we assume that all the gap junctions have the same conductivity value, i.e. $R_{i,j-1} = R_{i,j+1} =  R_{i-1,j} = R_{i+1,j} = R_{gap}$, we can simplify Eq. \ref{eq:grid_of_CMs_1} as 
\begin{equation}
\label{eq:grid_of_CMs_2}
\begin{split}
C_m\frac{dV_{i,j}}{dt}  
& = -I_{ion,i,j}  \\
& +\frac{V_{i,j-1}-2V_{i,j}+V_{i,j+1}+V_{i-1,j}-2V_{i,j}+V_{i+1,j}}{R_{gap}}\\
\end{split}
\end{equation}


\subsubsection{Whole heart modeling}
The natural expansion of the tissue modeling presented in the previous section is anatomically detailed 3D models of the electrical propagation in the whole heart or in the cardiac chambers (atrial and/or ventricular) \cite{cardone2016modeling3D, navarro2019modeling3D, rheeda2019modeling3D, ashikaga2013modeling3D, prakosa2018modeling3D, gillette2021modeling3D}. Although a detailed mathematical formulation is out of the scope of this review, we consider it useful for the reader to present the following pipeline that leads to these organ models. In case the organ model is not enveloped in a conductive medium, the monodomain equations (\ref{eq:monodomain}, \ref{eq:bidomain_bound_1}, \ref{eq:bidomain_bound_2}) are used \cite{sebastian2019modeling3D, rheeda2019modeling3D, ashikaga2013modeling3D, prakosa2018modeling3D}. Sebastian \textit{et al.} \cite{sebastian2009modeling3D} identified four main steps.
\begin{itemize}
    \item \textit{Segmentation of the cardiac geometry}. Firstly, it is necessary to build a realistic representation of the organ anatomy in order to shape the domain $\Omega$. Delayed enhancement Magnetic Resonance Imaging (DE-MRI) slices are generally used to build the 3D organ surfaces with the required level of detail, e.g., including papillary muscles and trabeculation, and labeling the model main components such as endocardium, epicardium, His bundle, and valve rings. Cardiac DE-MRI is an examination commonly done \textit{in vivo}, e.g., to assess infarcted areas, and it is suitable for obtaining patient-specific anatomically correct model surfaces \cite{minchole2019mri}.
    \item \textit{Volume meshing}. Secondly, in order to solve the monodomain reaction-diffusion problem, it is necessary to convert the 3D surface model into a 3D mesh. Traditionally, tetrahedral or hexahedral elements fill the surface model, resulting in a volumetric mesh.
    \item \textit{Myocardial fiber orientation}. As reported in the previous section, the conduction velocity is anisotropic, thus the propagation in the organ model is influenced by the fiber orientation. Such information can be acquired by diffusion tensor MRI (DT-MRI) and associated with each node in the volumetric mesh. For example, the Streeter \textit{et al.} method takes into account the helicoidal orientation of ventricular fibers \cite{streeter1969MRI}, while in papillary muscles and trabeculae, the orientation is aligned with the longitudinal axes. More advanced techniques, like the Laplace–Dirichlet rule-based algorithm by Bayer \textit{et al.} \cite{bayer2012FiberOrientation} allows the fiber orientation to change smoothly and continuously throughout the entire myocardium, also in models with complex geometries.  
    \item \textit{Fast conduction system}. Accurate simulations of the organ's electrical activation require a detailed conduction system model. Due to the difficulties in imaging the Purkinje cells, a common way to include the fast conduction systems is to consider scattered activation points or an estimation of tree-like structures from histology. In the latter case, 1D models, similar to the cable presented in the previous section, branch multiple times. In recent years, advanced human cardiac Purkinje cell models have been developed and are available to describe the detailed electrophysiology of the conduction system, e.g., \cite{trovato2020human}. 
\end{itemize}
However, the bidomain model enables more complex simulations, as the organ model can be enveloped in a conductive medium, e.g., as a torso model \cite{cardone2016modeling3D, navarro2019modeling3D, sebastian2019modeling3D}. Shortly, patient-specific torso models can be constructed by MRI slices, i) segmenting the lungs, bone regions, etc., ii) inserting the whole heart or biventricular model in the torso, and iii) finally meshing the torso. As the torso represents a passive domain, i.e., the reaction-diffusion problem is reduced to a diffusion problem, the torso meshing can be coarser than the heart meshing. We provide an overview of the applications of these anatomically detailed 3D models in the Discussion section.

\subsection{Brain models}

\begin{figure*}
    \centering
    \includegraphics[width=\textwidth]{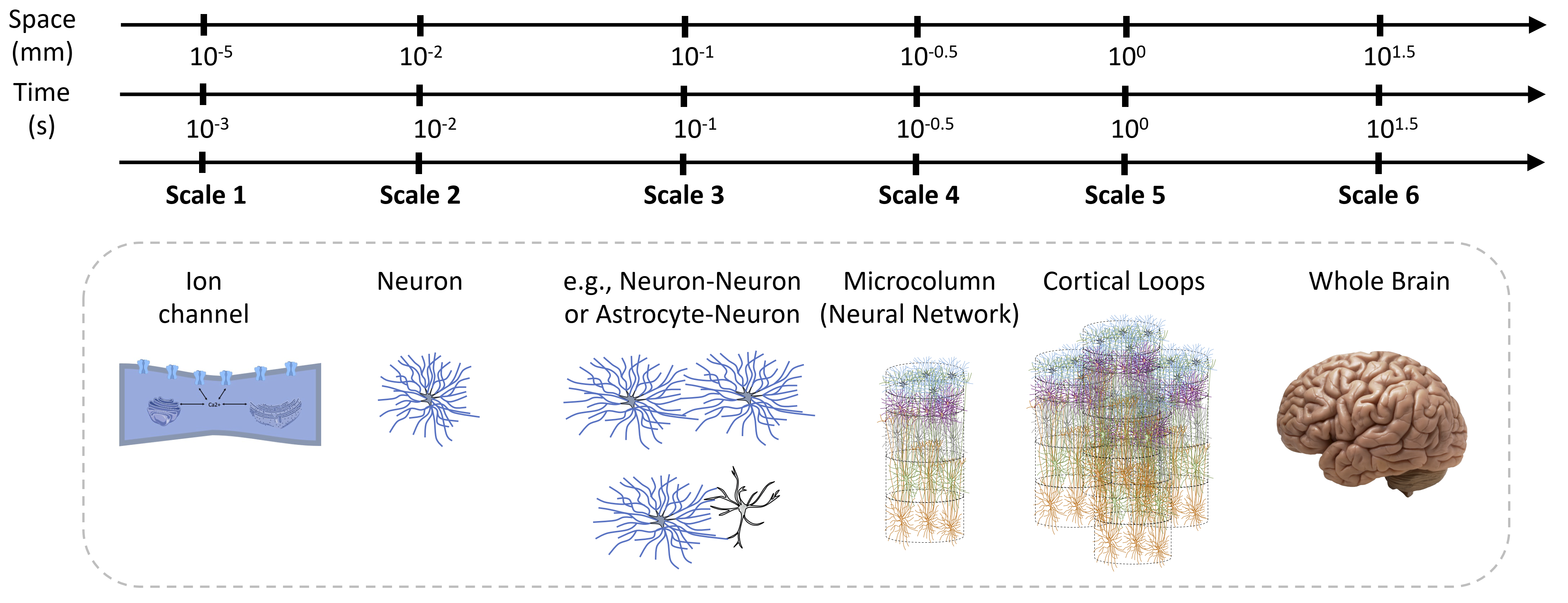}
    \caption{Six different spatial and temporal scales that are linked with the progression from intracellular pathways to organs in brain models.}
    \label{fig:rule_brain}
\end{figure*}

The rich dynamics of neural tissues inhibit the fast progress of its validated multiscale modeling. This results in a medium readiness level of a complete digital twin solution even though there are far more efforts in depicting the multiscale in brain tissues than any other type \cite{mattioni2013integration,martinolli2018multi}. The main challenge is the multiscale of the functional level and the plastic, complex, variable, and dense structure organization of the brain. Currently, the existing technology cannot be used to provide a complete understanding of it. At the functional level, we have the multiscale variability, depicted in Fig. \ref{fig:rule_brain} from ion channels \cite{faisal2008noise} to synapses \cite{destexhe2012neuronal}, single cells \cite{rabinovich2008transient,litwin2012slow,oschmann2018silico}, microcircuits or microcolumns \cite{boerlin2013predictive}, brain parts \cite{robinson2005multiscale,ferdousi2019nonlinear} and whole brain \cite{mizuno2010assessment,griffiths2019shaping}. However, there is a considerable gap between functional and organizational models, including the issue of having comprehensive models integrated into the multiscale spectrum \cite{sunkin2012allen}. Earlier, the focus was mostly on neuronal cells - now we are progressing to model multiple scales with other brain cell types such as glial cells like astrocytes, oligodendrocytes, and microglia.


\subsubsection{Neurons (scales 1 and 2)}

In \textit{Cable theory}, each neuron type can be modeled separately as a sequence of capacitances and resistances in parallel and then grouped to form an entire cortical microcircuit:
\begin{equation} \label{eq:cable_ion}
\tau \frac{\partial V_\eta}{\partial t} -
\lambda^2 \frac{\partial^2 V_\eta}{ {\partial x^2} }= -g_{l}(V_\eta-V_{l}) + I_{S} + I_{Na} + I_{Kd} + I_{M} + I_{T} + I_{L} 
\end{equation}
The \textit{Telegrapher's Equation} and the \textit{Compartment Models} are the most common mathematical frameworks to implement the cable theory (Eq. \ref{eq:cable_ion}), and it follows \cite{hodgkin1952simulation}, in which $\tau$ is the leakage conductance decay rate, $V$ is the membrane voltage, $x$ is the dendrite axis length, $\lambda$ is the spatial coordinate decay rate and $V_{L}$ is the leakage (or resting) potential of the cell. We also obtain the synaptic current $I_{S}$, which represents the total membrane voltage derived from a number of active terminals.  In the farthest right-hand side term, we have a summation of all ionic channels, which enriches the biological relevance of the membrane potential dynamics. Based on \cite{pospischil_minimal_2008},  $I_{Na}$ and $I_{Kd}$ are the sodium ($Na^+$) and potassium ($K^+$) currents responsible for action potentials, $I_{M}$ is a slow voltage-dependent $K^+$ current responsible for spike-frequency adaptation, $I_{T}$ is a high-threshold calcium (Ca$^{2+}$) current and $I_{L}$ is a low-threshold Ca$^{2+}$ current. The big advantage of this model is the rich number of ion channels, by which different types of neurons from each cortical layer can be replicated by fine adjustment of the ion channels possible. This is where experimental data provides fitted values to the variables in each ionic current description.

The voltage-dependent currents for the ion channels in the HH model can be obtained through variations of the same generic equation for a current $I_j$ defined  as

\begin{equation}
I_j = g_j m^M h^N (V_\eta-E_j),
\end{equation}

\noindent where $g_j$ is the maximal conductance. The parameters $m$ and $h$ are the activation and inactivation variables respectively with the order of $M$ and $N$, followed by the difference between the membrane potential $V_\eta$ and the reversal potential $E_j$. The steady-state activation and time constants are, respectively, given by $m_\infty 0 \alpha/(\alpha + \beta)$ and $\tau_m = 1/(\alpha+\beta)$. Similarly, it is for $h$, in which $\alpha$ is the conditional rate of active ion channel based on $V_\eta$ and $\beta$ is the conditional rate of inactive ion channel based on $V_\eta$.

\subsubsection{Neuron-astrocyte interactions (scale 3 and 4) }

\begin{figure}
    \centering
    \includegraphics[width=0.55\textwidth]{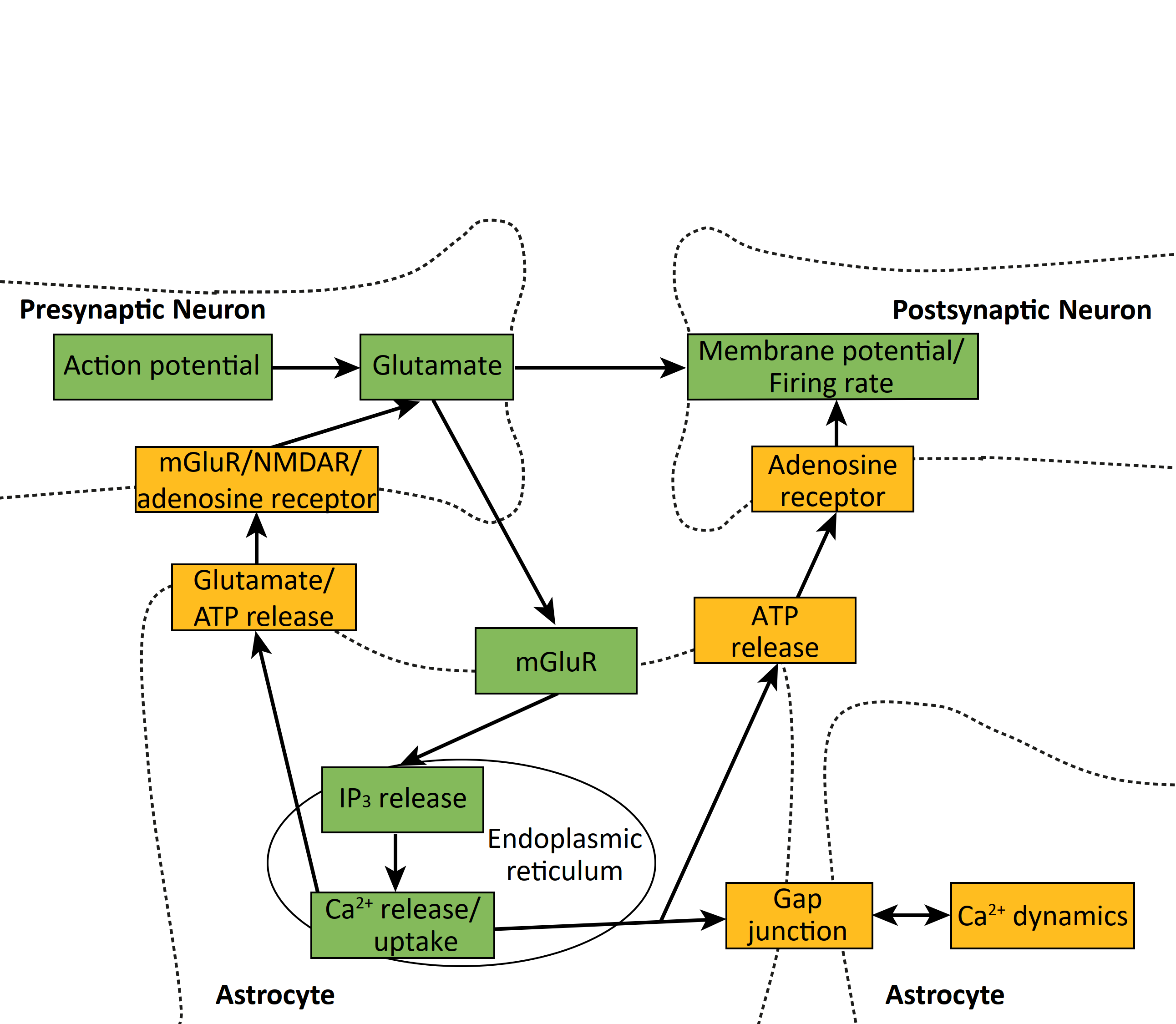}
    \caption{Tripartite synapse model including a pre- and a postsynapse contacted by an astrocyte. An action potential in the presynapse triggers glutamate release which can be taken up by the postsynapse. Glutamate can also bind to the metabotropic glutamate receptors (mGluR) of the astrocyte. Then, a cascade of IP$_3$ and Ca$^{2+}$ release from the endoplasmic reticulum follows and may induce a Ca$^{2+}$-induced Ca$^{2+}$ release. Mainly IP$_3$ diffuses through gap junctions to the neighboring astrocytes. Higher Ca$^{2+}$ concentrations within the astrocyte can trigger a transmitter release of glutamate and ATP/ adenosine towards the neurons.  }
    \label{fig:tripartSynapse}
\end{figure}

Nadkarni and Jung \cite{Nadkarni2007} introduced a tripartite synapse model, meaning that they included a pre- and a postsynaptic neuron and an adjacent astrocyte (Figure~\ref{fig:tripartSynapse}). The presynaptic pyramidal neuron is formulated as a Pinsky-Rinzel model \cite{pinsky1994intrinsic} with two compartments: 

\begin{equation}
    \begin{split}
C_{m} \frac{\mathrm{d} V_{s}}{\mathrm{~d} t} &=-I_{\mathrm{Leak}}\left(V_{s}\right)-I_{\mathrm{Na}}\left(V_{s}, h\right)-I_{\mathrm{K}-\mathrm{DR}}\left(V_{s}, n\right) \\
&\quad+\frac{g_{c}}{p}\left(V_{d}-V_{s}\right)+\frac{I_{s}}{p} \\
C_{m} \frac{\mathrm{d} V_{d}}{\mathrm{~d} t}& =-I_{\mathrm{Leak}}\left(V_{d}\right)-I_{\mathrm{Ca}}\left(V_{d}, s\right)-I_{\mathrm{K}-\mathrm{AHP}}\left(V_{d}, w\right) \\
&\quad -I_{\mathrm{K}-\mathrm{C}}\left(V_{d},\left[\mathrm{Ca}_{\mathrm{neuron}}\right], c\right)+\frac{g_{c}}{1-p}\left(V_{s}-V_{d}\right)\\
&\quad+\frac{I_{d}}{1-p}.
\end{split}
\end{equation}

The parameters $V_s$ and $V_d$ denote the somatic and dendritic membrane potentials, which are measured in relation to a reference potential of $-60~mV$. $I_d$ is the  current injected into the dendrite divided by the total membrane area. The parameter $p$ represents the fraction of the somatic cell volume. The following currents are included:  leakage ($I_{leak}$), $Na^+$ ($I_{Na}$), delayed rectified $K^+$ ($I_{K-DR}$), and Ca$^{2+}$ ($I_{Ca}$). 


The glutamate contained in a vesicle released from the presynapse binds to the AMPA ($\alpha$-amino-3-hydroxy-5-methyl-4-isoxazolepropionic acid) receptors at the postsynapse:

\begin{equation}
\begin{split}
I_{\mathrm{AMPA}}&=g_{\mathrm{AMPA}} Y_{\mathrm{AMPA}}\left(V_{\mathrm{inh}}-V_{\mathrm{syn}}\right) \\
\frac{\mathrm{d} Y_{\mathrm{AMPA}}}{\mathrm{d} t}&=\Theta\left(t-t_{\mathrm{rel}}\right)-\Theta\left(t-t_{\mathrm{rel}}-1 \mathrm{~ms}\right)-\frac{Y_{\mathrm{AMPA}}}{1.0}.
\end{split}
\end{equation}

The maximum conductance is given by $g_{AMPA}$ and the voltage $V_{syn}$ by the reversal potential of the $Na^+$ conductance of the postsynaptic interneuron, $E_{Na} = 55~mV$. $Y_{AMPA}$ denotes the stochastic first-order kinetics of the AMPA receptor and a Heaviside function for the initiation of the postsynaptic current. The membrane potential of the interneuron is then modeled as follows:

\begin{equation}
C_{\text {inter }} \frac{\mathrm{d} V_{\text {inter }}}{\mathrm{d} t}=-I_{\mathrm{Na}, \text { inter }}-I_{\mathrm{K}, \text { inter }}-I_{\mathrm{L}, \text { inter }}+I_{\mathrm{AMPA}},
\end{equation}

where $I_{\mathrm{Na}, \text { inter }}$, $I_{\mathrm{K}, \text { inter }}$ and $I_{\mathrm{L}, \text { inter }}$ are the respective $Na^+$, $K^+$ and leak currents.

The glutamate released from the presynapse also binds to metabotropic glutamate receptors (mGluR) at the membrane of a nearby astrocyte (Figure~\ref{fig:tripartSynapse}). The binding leads to a release of inositol trisphosphate (IP$_3$) from the endoplasmic reticulum (ER) and subsequently to Ca$^{2+}$ release. The intracellular Ca$^{2+}$ concentration is described as:

\begin{equation}
\frac{\mathrm{d}\left[\mathrm{Ca}_{\text {astro }}^{2+}\right]}{\mathrm{d} t} =-J_{\text {Channel }}(q)-J_{\text {Pump }}-J_{\text {Leak }}, 
\end{equation}
where $J_{Channel}$ is the flux from the ER to the cytosol through the fraction of open IP$_3$ receptor channels $q$, $J_{Pump}$ the flux from the cytosol to the ER and $J_{Leak}$ the leak flux from the ER to the cytosol. This simplified model of an astrocyte follows the Hodgkin–Huxley model \cite{hodgkin1952simulation} by exchanging the transmembrane potential with the Ca$^2+$ concentration.

Because of the complexity of generating biophysical models at this particular scale and above, models tend to become more phenomenological instead of pure biophysical. While many argue that a phenomenological model is a black box, we also counter-argue that a phenomenological model in a multiscale scenario with underlying biophysical models supporting a certain phenomenological model brings its black box approach to a more gray box one. 

Lenk \textit{et al.}~\cite{Lenk2020} introduced a phenomenological neuron-astrocyte network model called INEXA, which contains both excitatory and inhibitory neurons. The model was initially developed to study the astrocytes' effect on neuronal firing rates in the cortex. The neurons and astrocytes in the network are connected via tripartite synapses. The governing equation for the  firing rate $\lambda _i$ of a postsynaptic neuron $i$ for each time step $t_k$ of $5$ $ms$ is:
\begin{equation}
 \lambda_i (t_k)  =  \max \left(0, c_i + \sum_j y_{ij} \cdot s_j (t_{k-1}) 
- \sum_j y_{\text{Astro}} \cdot A_{ija} (t_{k-1}) \right).\\
\end{equation}
The parameter $c_i$ is denoted as stochastic noise of neuron $i$. The term $y_{ij}$ represents the synaptic weight between the presynaptic neuron $j$ and a postsynaptic neuron $i$, which can be either excitatory ($y_{ij} \in [0,1]$, corresponding to glutamatergic neurons) or inhibitory ($y_{ij} \in [-1,0]$, corresponding to GABAergic interneurons). The parameter $s_j$ indicates whether neuron $j$ transmitted a spike in the previous time step $t_{k-1}$. The second part of the equation denotes the depressing effect caused by astrocytes on the excitatory synapses. The variable $A_{ija}$ shows whether the astrocyte \textit{\texttt{a}} connects to  the synapse $ij$  and if astrocyte \textit{\texttt{a}} was activated at the previous time step. 

Glutamate is released by excitatory synapses into the synaptic cleft with rate $\Omega_f$, which binds to astrocytic mGluRs with rate $\Omega_g$. This causes an IP$_3$-mediated release of Ca$^{2+}$ from the ER into the astrocytic cytoplasm (Figure~\ref{fig:tripartSynapse}). The governing equation for astrocyte dynamics is based on the intracellular Ca$^{2+}$ dynamics $[Ca^{2+}]_{ija}$:
\begin{equation}
[Ca^{2+}]_{ija}(t_k) =  [Ca^{2+}]_{ija}(t_{k-1}) 
+ \Omega_{acc} \cdot ([IP_3]_{ija}(t_k)  - [Ca^{2+}]_{ija}(t_{k-1})).
\end{equation}

The intracellular Ca$^{2+}$ concentration consists of the Ca$^{2+}$ concentration left from the last time step ($Ca^{2+}_{ija}(t_{k-1})$), the IP$_3$-mediated Ca$^{2+}$-induced Ca$^{2+}$-release from the ER stores, and the Ca$^{2+}$ uptake back to the ER by the SERCA (sarco/endoplasmic reticulum Ca$^{2+}$-ATPase) pumps. The slow dynamics of the Ca$^{2+}$ release are considered by a multiplication of the ER term with the time scale $\Omega_{acc}$. 

The study~\cite{Lenk2020,lenk2021larger} showed that astrocytes modulate neuronal bursting and can prevent hyper-excitability of neurons. In Fritschi \textit{et al.}~\cite{Fritschi2021}, the INEXA model was used to test several hypotheses of the astrocytic impact on schizophrenia. 

\subsubsection{Microcircuits/ Microcolumns (scale 4 and 5)}


\begin{figure}
\centering
\includegraphics[width=.65\textwidth]{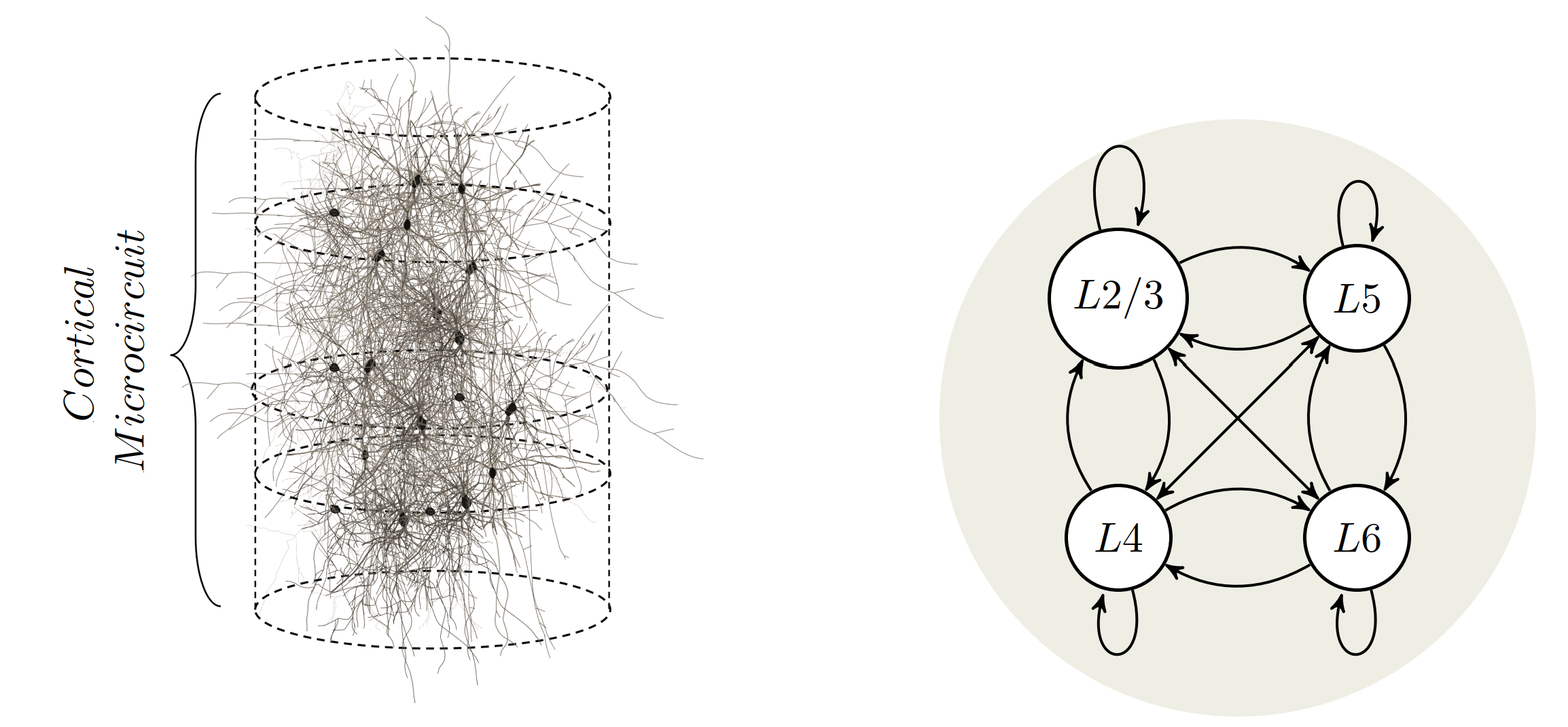}
    \caption{(a) Cortical circuit presenting different physiological characteristics, such as connection, hierarchy, and morphology that ultimately impact the channel capacity. (b) The cortical layer hierarchy can be captured by the relationship between connection probabilities between cortical layers that has been extensively explored \textit{in vitro} and modeled as a Markov chain. The image presents the Markov chain model used in this paper, with both pre and post-synaptic connection probabilities. 
    }
\label{fig:markov_chain_wprobs} 
\end{figure}

Moving up on the spatiotemporal scales, the brain exhibits the structure of biological neural networks with different cell types in different brain locations, which can be termed microcircuits. We also consider that this scale should incorporate groups of biological networks that form a more complex type of microcircuits that are, however, mainly located in the gray brain areas and are termed microcolumns. Due to its increased complexity with even more cell types being considered, microcolumns are also divided into layers with which individual cells are associated with. 
The somatosensory cortical layer hierarchy can be captured by the relationship of connection probabilities between cortical layers that have been explored \textit{in vitro} \cite{ramaswamy2015neocortical,adonias2021analysis}. This area has been the main focus of efforts for the digitalization of the human brain, including the Blue Brain Project. They have considered a phenomenological approach where cell connections are depicted in a probabilistic manner. Even though this approach helps to characterize a connectome of some of the brain areas, the true complexity and characterization should incorporate the dynamics at scales that describe the true biophysical phenomenon \cite{adonias2020utilizing}. However, since this approach has not been yet explored in the literature, we will depict a phenomenological model in the following.

The basic idea is to provide a description of the network connections and time-variant connection changes in a probabilistic manner based on what has been observed in existing microcircuit modeling and digital reconstruction efforts \cite{barros2018capacity,ramaswamy2015neocortical}. Many probabilistic tools that describe network connections can be employed. However, we believe Markov Chains enable us to algebraically investigate the many neuronal connections and their plastic behavior at the macroscale using identity matrices and corresponding matrix operations (eigenvalues, eigenfunctions, etc). Consider a \textit{discrete-time} Markov chain with $|\mathbf{N}|$ states, which is also the number of cortical layers, and with $\mathbf{N} = \{2/3,4,5,6\}$ states, which correspond to layers 2/3, 4, 5 and 6 respectively. The state $2/3$ is defined for cortical layers 2 and 3 simultaneously. The transition probability matrix $\mathbf{P}$ is characterized by $|\mathbf{N}| \times |\mathbf{N}|$ elements, where $\mathbf{P}$ should satisfy 

\begin{equation}
\forall i,j,P_{i,j} \in [0,1],
\end{equation}

\noindent and, 

\begin{equation}
\forall i, \sum_{j=1}^{|\mathbf{N}|} P_{i,j} = 1. 
\end{equation}

To compute $\mathbf{P}$, we consider two different Markov chains that account for the \textit{presynaptic connection probabilities} ($\mathbf{N_\alpha}$) and the \textit{postsynaptic connection probabilities} ($\mathbf{N_\beta}$), with transition probabilities matrices ($\mathbf{P_\alpha}$) and ($\mathbf{P_\beta}$), respectively. The ($\mathbf{P_\alpha}$) and ($\mathbf{P_\beta}$) initial probabilities were obtained from \cite{ramaswamy2015neocortical} and are depicted in Fig.~\ref{fig:markov_chain_wprobs}. Finally, considering independent connections, we can represent the post- and presynaptic connections as 

\begin{equation}
\mathbf{P} = \mbox{E}[\mathbf{P_\alpha} + \mathbf{P_\beta}],
\end{equation}

\noindent in which $E[\cdot]$ is the expected value. For simplicity, we consider that the network model is defined by a binary adjacent matrix $\mathbf{\Upsilon}$ of $\mathbf{N}$ and $\mathbf{\Upsilon}(\mathbf{N}) = \mathbf{\Upsilon}(\mathbf{N_\alpha}) = \mathbf{\Upsilon}(\mathbf{N_\beta})$. 


We model the connectivity of a microcircuit using a transition stochastic matrix of an ergodic Markov Chain. Such a stochastic matrix can be defined by $P_{i,j}$, as the $i$-th row and $j$-th column element. Such $i$ and $j$ represent the non-ordered set of neurons as part of nodes of the network. 

\begin{equation}
    P=\left[{\begin{matrix}P_{1,1}&P_{1,2}&\dots &P_{1,j}&\dots &P_{1,S}\\P_{2,1}&P_{2,2}&\dots &P_{2,j}&\dots &P_{2,S}\\\vdots &\vdots &\ddots &\vdots &\ddots &\vdots \\P_{i,1}&P_{i,2}&\dots &P_{i,j}&\dots &P_{i,S}\\\vdots &\vdots &\ddots &\vdots &\ddots &\vdots \\P_{S,1}&P_{S,2}&\dots &P_{S,j}&\dots &P_{S,S}\\\end{matrix}}\right]
\end{equation}

We then need to follow the property of

\begin{equation}
    \sum _{j=1}^{S}P_{i,j}=1,
\end{equation}

\noindent where $S$ is the cardinality of the matrix.

Since plasticity in a phenomenological model can be reduced to a probability term, we hypothesize that a plasticity model for a microcircuit is a time-dependent matrix that directly changes the stochastic transition matrix. Therefore, we can incorporate any existing biophysical plasticity model into a phenomenological model, providing an interesting option for multiscale modeling. We can define such plasticity term ($\mathbf{Pl}(t)$) as 

\begin{equation}
\mathbf{Pl}(t) = f*(t)\mathbf{P}, \quad t\in [0,\infty),
\end{equation}

\noindent where $f*(t)$ is a normalized plasticity biophysical function and must be $f*(t) \neq 0$.

Many approaches to obtaining a plasticity biophysical function $f(t)$ exist, depending on the level of molecular interactions one wants to study. The most conventional rate-based models or spike timing-based models offer less complex approaches to account for the effects of spike generation at the post-synaptic neurons \cite{andrade2023timing}. Even though they can elicit easier integration with multi-scale modeling, this approach is mostly used for understanding higher-level phenomena, e.g., receptive fields, learning, and visual cortical layering. For a better understanding of how synaptic plasticity relates to biophysical modeling, we can thus consider models of Ca$^{2+}$-dependent or signal transduction-dependent synaptic plasticity, explained in the following. First, let's define $W_i$ as the synaptic weight of the $i$-th synapse. This synaptic weight is the same as previously described for scales 1 and 2. Now, considering

\begin{itemize}
    \item \textbf{Ca$^{2+}$-dependent synaptic plasticity}: The synaptic weight is defined by relating with the Ca$^{2+}$ transient behavior, which is associated with long-term depression and long-term potentiation and thus with the connection changes due to the behavior of postsynaptic neuron response to neurotransmitter release. More directly, Ca$^{2+}$ transients coordinate the neurotransmitter release and production. The following model then, proposed by \cite{karmarkar2002model}, introduces Ca$^{2+}$ as the central variable in the changing of $W_i$ over time, as
    
    \begin{equation}
       f(Ca) = {\frac{dW_i}{dt}}=\eta(Ca)\left(\Omega(Ca)-\lambda W_i\right)
    \end{equation}
    \noindent where $\Omega$ is the Ca$^{2+}$ transient magnitude, $\eta$ is the Ca$^{2+}$-dependent, and $\lambda$ is a decay constant.
    
    \item \textbf{Signal transduction-dependent synaptic plasticity}: The synaptic weight is based on the dynamical assessment of the potentiation and depression phenomena. It is a biophysical model but with a 	tendency to a phenomenological model since it tries to abstract the molecular level layer with a higher level modeling between a potentiation and depression function. These functions will characterize the dynamics involved in the potentiation and depression events, and correlate to some of the transduction mechanisms that dictate synaptic behavior.  We can thus define that the synaptic weight depends on a potentiation ($P(t)$) and a depression ($D(t)$) function as
    \begin{equation}
        f(g_0,\gamma){\frac{dW_i}{dt}}=g_0\{P(t)D(t)^\gamma-D(t)P(t)^\gamma\},
    \end{equation}
    \noindent where $g_0$ is a scaling constant and $\gamma$ the competition degree that dictates the potentiation and depression functions.
\end{itemize}



\subsubsection{Cortical Loops (scale 5)}

Modeling beyond microcircuits requires an alternative approach since the compartmentalization method for the various Hodgkin-Huxley functions requires extreme computing resources in order to host such digital twins. The EBRAINs (\url{https://www.ebrains.eu/}) platform for the EU Human Brain project exemplifies such requirements - a data center-like infrastructure is required to run around 1mm$^3$ of microcolumn tissue, which comprises roughly 10,000 neurons of different types. As they concentrate on the neocortex, however, other works provide evidence of a similar setup for the hippocampus as well as the thalamus \cite{tome2022coordinated}.

The literature explores a mathematical formulation that simplifies the description of action potential propagation in populations or multiple populations, using a simple set of equations. The major challenge is that the degree of freedom in this approach is higher, and more likely to produce erroneous results since many of the dynamics/non-linearities are simplified. However, with a precise definition of a specific type of activity and a precise focus on parts of the brain, this type of approach can be quite powerful. 

Two different modeling approaches can be used for simulating population-level activity including mass models or mean-field equations. Both have been demonstrated in the literature to work well with empirical data and validate events of self-sustained local oscillations and patterns of distant synchronization \cite{deschle2021validity}. Even though they have the same goal, their differences can favor different types of studies. While mass models can afford the addition of new elements towards dynamism explanation \cite{deschle2021validity,byrne2020next}, mean-field models concentrate on simpler and close-to-local behaviors while maintaining fidelity to empirical data. From this paper's point of view, both present reasonable valid approaches to modeling cortical loops. Here, we focus on already explored thalamus-cortex cortical loops.

\begin{figure}
    \centering
    \includegraphics[width=.45\textwidth]{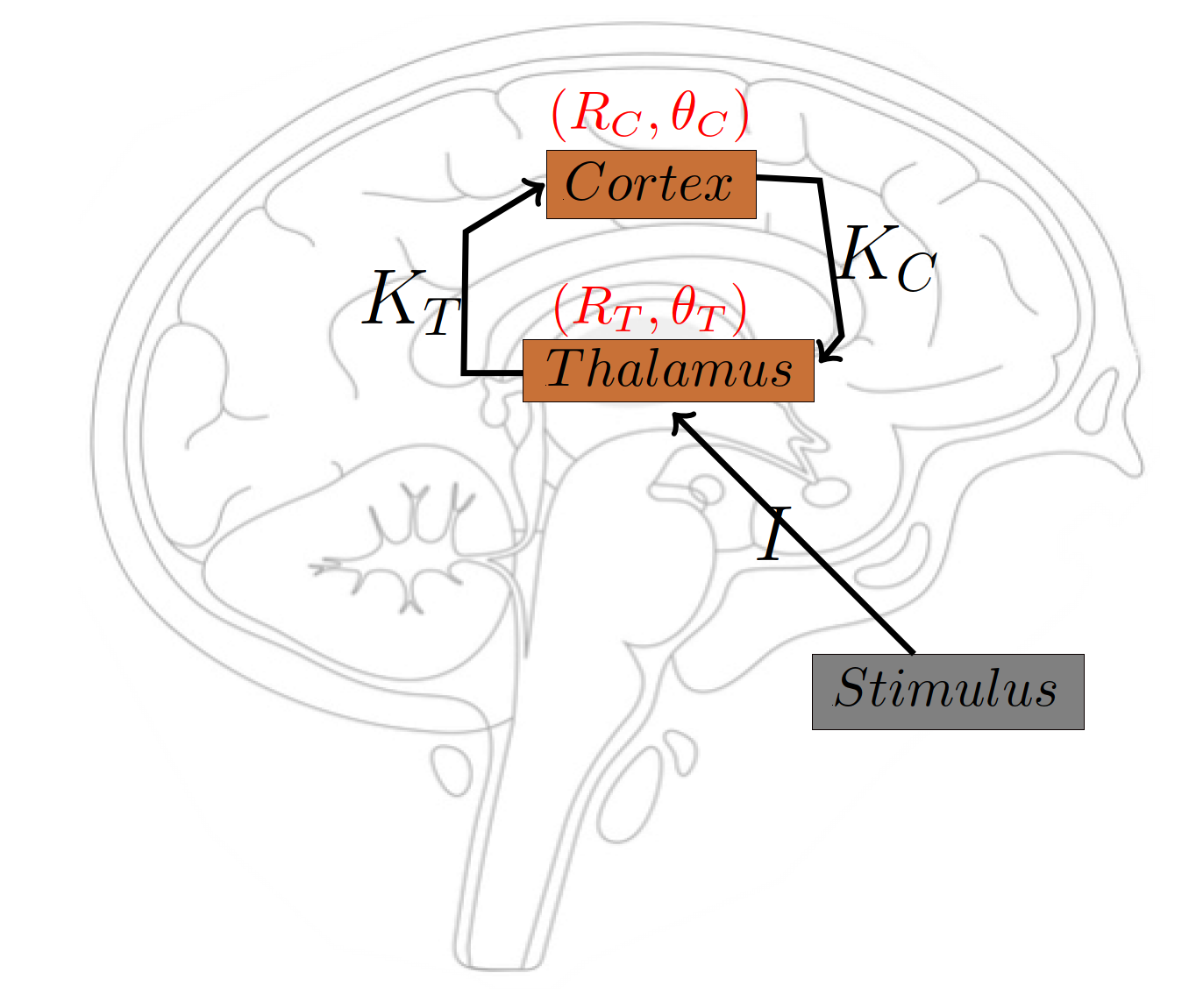}

    \caption{Modelling of a cortical loop between thalamus and cortical populations following a mean-field approach based on a given stimulus. The relationship of the population is used to understand $\theta$-band oscillations.}
    \label{fig:cortical_loops}
\end{figure}


Rosjat \textit{et al}. \cite{rosjat2014mathematical} presented a minimal model of the thalamus and the cortex using complex mean-field approaches. Each population comprises neural oscillators aiming to describe the brain wave activity in both the thalamus and cortex within the $\theta$-band spectrum. Here, the mean fields represent a coarse-grained description of their spatial arrangement as a population and how the two populations relate to each other in communicating electrophysiological activity. This is another level of phenomenological modeling since the population description is a cumulative value compared to the detailed electrophysiological tracking model presented before. The reason for this is that areas of the hypothalamus or the cortex will now present millions of neurons, and therefore, even probabilistic models, with reduced complexity, will not be able to present valuable computationally feasible solutions. The description of the two populations using the complex mean field is 

\begin{equation}
\begin{array}{l}
\frac{d \phi_{T}\left(\omega_{T}\right)}{d t}=\omega_{T}+K_{C} R_{C} \sin \left(\theta_{C}-\phi_{T}\left(\omega_{T}\right)\right)\\+I(t) \cos \left(\phi_{T}\left(\omega_{T}\right)\right)\\
\frac{d \phi_{C}\left(\omega_{C}\right)}{d t}=\omega_{C}+K_{T} R_{T} \sin \left(\theta_{T}-\phi_{C}\left(\omega_{C}\right)\right) \end{array},
\end{equation}

\noindent where $\omega_T$ and $\omega_C$ are continuous variables distributed in each population of oscillators as

\begin{equation}
n_{a}\left(\omega_{a}\right)=\frac{2}{\pi\left(1+4\left(\omega_{a}-\tilde{\omega}_{a}\right)^{2}\right)}, \quad a=\{T, C\}
\end{equation}

\noindent and represent the natural frequencies of these oscillators. $\phi_{T}$ and $\phi_{C}$ are the phases of these oscillators of the thalamus and cortex populations, respectively. The state of each population can be described by the distribution density $W (x, \phi, t) = n(x)w(x, \phi, t)$, with the conditional distribution density of oscillators denoted by $w(x, \phi, t)$. These probability density distributions can be either estimated arbitrarily or from actual experimental data, as the latter is the best option for pre-validation.

Each oscillator in the cortical population is coupled to the complex mean field

\begin{equation}
Y_{T}=R_{T} e^{i \theta_{T}}=\int n(x) \int_{-\pi}^{\pi} e^{i \phi_{T}} w\left(x, \phi_{T}, t\right) d \phi d x
\end{equation}

\noindent of the thalamic population and each oscillator in the thalamic population is coupled to the complex mean field

\begin{equation}
Y_{C}=R_{C} e^{i \theta_{C}}=\int n(x) \int_{-\pi}^{\pi} e^{i \phi_{C}} w\left(x, \phi_{C}, t\right) d \phi d x
\end{equation}

\noindent of the cortical populations (see Fig. \ref{fig:cortical_loops}). Coupling strengths are $K_T$ and $K_C$, respectively.

The thalamus population is stimulated by an external stimulus that acts directly on it (see Fig. \ref{fig:cortical_loops}). The stimulus term is basically the signal entry point from other areas of the brain. However, it can be used for external signal inputs from stimulation devices, even though this has not been exactly in the referred model. This stimulus is represented by the term $I(t)$cos$(\phi T)$ where

\begin{equation}
I(t)=\left\{\begin{array}{ll}I & \text { during stimulus, } \\ 0 & \text { otherwise. }\end{array}\right. 
\end{equation}

Mean-field dynamic modeling such as presented here should be carefully integrated with lower-scale models and incorporate neuronal activity types such as populations that integrate both excitatory and inhibitory neurons. The further the level of dynamics in the models is, the more connections it needs to lower scales in order to link the source of dynamical behavior and mean-field approximations. This will help the credibility of these more gray models. Issues such as population synchronization can then be tackled when more complete models emerge since variations of spike rate convoluted to operating synchronization frequency can then be mathematically described.

\subsubsection{Whole brain (scale 6)}


A whole brain model is clearly the most challenging as many dynamics of different brain parts hinder the agreement on an actual methodology that is validated (even if in parts), personalized (brings different structural and functional changes based on the subject's brain in question), and in terms of model complexity. In this section, we focus on the most recent methodology that has been proposed by Kringelbach \textit{et al.} \cite{kringelbach2020dynamic}. The main advantage is that brain imaging is coupled with temporal functional modeling which allows us to address many past challenges. First, a multiscale model can be broken down into the following main parts:

\begin{itemize}
    \item \textbf{Neurodynamical system} comprising of spontaneous brain activity at the level of the whole brain where each node in the network represents a brain region and the links between nodes represent white matter connections.
    \item \textbf{Blood oxygen level-dependent (BOLD) signal} which is the relation of the population level activity to the homeodynamic state of that tissue given by the deoxyhemoglobin content from the extravascular and intravascular signals.
    \item \textbf{Parcellated structural and functional data} from cortical and subcortical brain regions.
\end{itemize}

Kringelbach \textit{et al.} \cite{kringelbach2020dynamic} termed the approach \textit{dynamic mean-field (DMF)}. As with other mean-field approaches, the model is based on excitatory and inhibitory pools of neurons which are coupled with a guiding factor. Their coupling factor takes into account the inhibitory and excitatory currents, the GABA and NMDA receptors, however, only with excitatory-to-excitatory connections. Some more recent work has taken another step to propose a \textit{vultiscale dynamic mean-field (MDMF)}, where the microdomain kinetics of the synaptic cleft are used for the coupling of the neuronal populations \cite{naskar2021multiscale}.

\subsection{Epithelial models}

\begin{figure*}
    \centering
    \includegraphics[width=\textwidth]{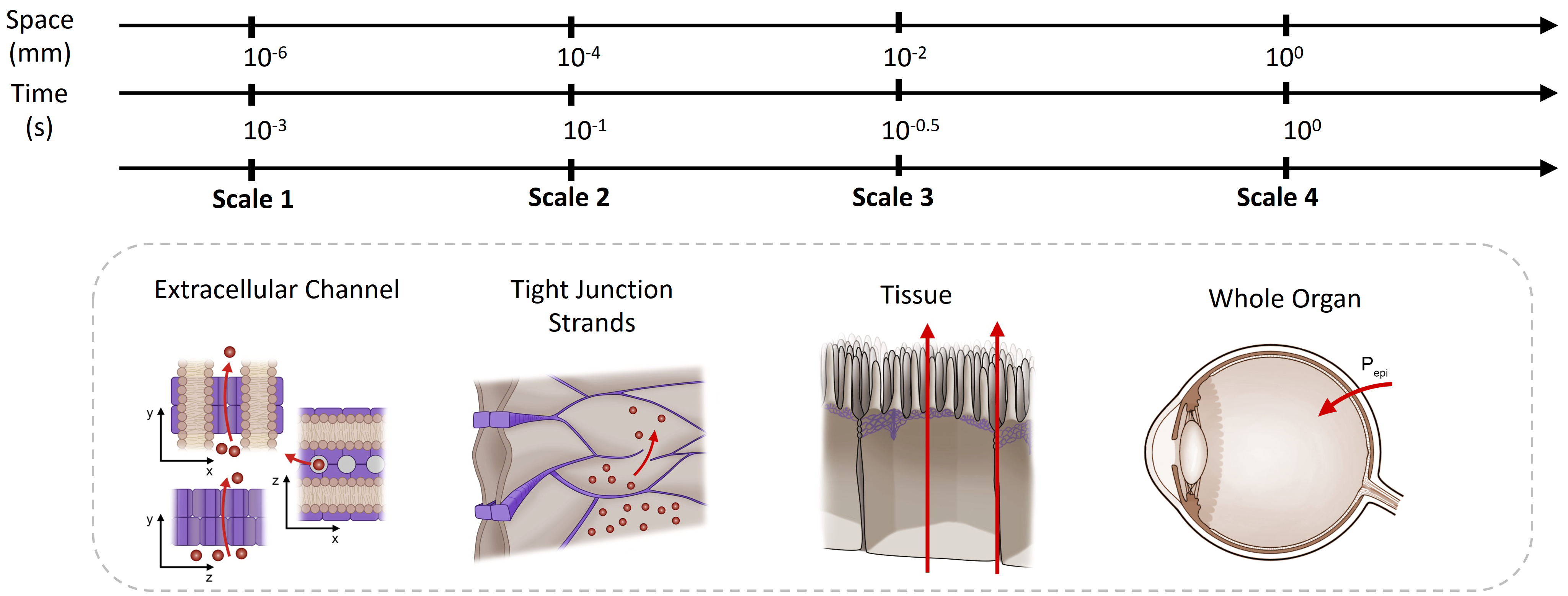}
    \caption{Different spatial and temporal scales that are linked with the progression from molecular diffusion to organs with epithelial tissues.}
    \label{fig:rule_ep}
\end{figure*}

Modeling epithelial tissue with multiscale approaches can lead to developments in the understanding of a variety of pathologies originating from disorders in the epithelial barrier. Furthermore, the role of epithelia as the barrier against drug delivery makes them an important component in drug delivery to any tissue. The bulk of the computational research on epithelia has focused on embryo- and organogenesis. The models of the epithelial barrier are mostly limited to the scale of the tissues. Even though epithelial tissues are everywhere in the body and their functions are integrated into most organ systems, there have been minimal efforts on multiscale organ-level models. There is also a great variety of diseases and conditions that can highlight the utility of digital twins for epithelial systems. In this section, we will focus on epithelial barriers and detail how their digital twins can be constructed and how epithelial multiscale influences their organ-specific functionalities. Unlike cardiac and neuronal cells, the epithelial tissues often support multiple functions specific to the adjacent tissues (mostly related to organ-specific tight junctions and barriers).

The epithelial barrier is usually divided into two components: the transcellular (through the cells) and the paracellular (between the cells). The majority of the transcellular component is formed by the cell plasma membrane and its specific channels and transporters. On the other hand, the governing paracellular barrier component is formed by the  tight junctions, molecular structures that close off the space between the cells. These semipermeable structures are formed by intercellular connections that organize into a 2D network of strands between neighboring cells \cite{vanitallie2014,otani2020}. The barrier properties are generally studied by measuring the permeability of various neutral molecules or the transepithelial electrical resistance (TER). Small molecules and ions can pass through the tight junctions via extracellular channels formed by claudins, one of the main structural transmembrane proteins in these junctions \cite{watson2001,vanitallie2008}. However, also larger molecules can pass via the so-called leak pathway, whose accurate origin is unclear.

\subsubsection{Claudin channels (scale 1)}

The extracellular/paracellular channels  (Fig.~\ref{fig:rule_ep}), formed by some claudin proteins, enable the movement of ions and small molecules from one side of epithelium to the other, rather than between the intra- and extracellular space. The channels form charge- and size-selective pores and have been shown to be dynamically gated similar to ion channels. Weber \& Turner \cite{weber2017} modeled the conductivity of the stochastically gated claudin-2 channels and described them as having two closed states, stable ($C1$) and unstable ($C2$), and an open state ($O$). The states changed based on the following transition probability matrix, with the transition from $C1$ to $C2$ omitted: 

\begin{equation}
    Prob = \left[\begin{matrix}  
p_{O,O} & p_{O,C1} & p_{O,C2}\\  
p_{C1,O} & p_{C1,C1} & 0\\  
p_{C2,O} & p_{C2,C1} & p_{C2,C2}
\end{matrix}\right],
\end{equation}
where $p_{i,j}$ is the probability of transiting from state $i$ to state $j$ and $\sum_j p_{i,j} = 1$. The channel conductivity could then be defined based on their state and the conductivity of each state.

The molecular permeability of these channels has been modeled using the so-called \textit{Renkin function} \cite{guo2003,vanitallie2011,avdeef2010,dechadilok2006}, where the channel is described by a circular pore with the permeability calculated as

\begin{equation}
P_{pore} = \frac{A_{pore} D_{free} F\left(r_m/R_{pore}\right)}{d_{pore}}.
\label{eq:pore}
\end{equation}
The parameter $A_{pore}$ is the total pore area per unit tissue area, $D_{free}$ is the molecule’s free diffusion coefficient, $d_{pore}$ is the pore depth, and $F\left(r_m/R_{pore}\right)$ is a Renkin-type hindrance as a function of the relative size between the radius of the molecule, $r_m$, and the  pore radius $R_{pore}$. Different formulations for the hindrance function $F$ have been compared in  \cite{dechadilok2006}.

The claudin pores have also been the target of molecular scale diffusion studies \cite{yu2008} and molecular dynamics simulations, with the aim to identify the charge-selective component in the channels \cite{samanta2018,alberini2017}. However, the studies are outside the scope of this review.

\subsubsection{Tight junction strand network (scale 2)}

Tight junctions have also been modeled in the scale of the strands and the network they form (Fig. \ref{fig:rule_ep}). Here, the focus has been on the different pathways through these junctions as well as on the effect of compartmentalization and strand-level dynamics. Weber \& Turner \cite{weber2017} simulated the effect of the stochastic claudin-2 gating behavior – in combination with the compartmentalization in the strand network – on the resulting TER. In essence, they defined a circuit with stochastic resistors that form in the strand network and solved the time-dependent linear system of equations to replicate their tight junction patch-clamp experiments.

Different components have been proposed to form the leak pathway in the tight junctions, including structural strand dynamics manifested by constant strand breaking and annealing, and large tubes in the tricellular junctions at the meeting points of three cells. Guo \textit{et al.} \cite{guo2003} modeled the tight junctions by a static two-pathway strand model composed of multiple small pores and large rare breaks. They described the two pathways in parallel with differing contributions for small molecules and ions. 

The model by Tervonen \textit{et al.} \cite{tervonen2019} described these rare breaks as dynamic phenomena to represent the leak pathway, in combination with the large static tubes at the tricellular junctions. The claudin-channel dynamics were omitted in the model, as it is concentrated on the formation of the leak pathway. Here, the model was built to describe both the molecular permeability and TER measurements, and the strand dynamics were described by stochastic rate constants or resistors, respectively, between the tight junction compartments.

The permeability model was based on solving the amount of substance in each compartment by using the following ordinary differential equation:

\begin{equation}
\frac{d q_i (t)}{dt} = \sum_{i\neq j}^n \left( k_{ij}(t)q_j(t) - k_{ij}(t)q_i(t)\right),
\end{equation}
where $q_{i}$ is the amount of substance in compartment $i$, and $k_{ij}$ is the stochastic rate constant between compartments $i$ and $j$, whose value was calculated based on a stochastic variable to describe the breaking and annealing behavior. The final permeability of this system can then be calculated from the apical concentration as a function of time as

\begin{equation}
P_{network} = \frac{dq_{apical} (t)}{dt} \frac{1}{w_{model} c_{basal}},
\end{equation}
where $w_{model}$ is the width of the modelled section and $c_{basal}$ is the basal concentration. The static tricellular tube permeability can be calculated using an equation of a type shown in Eq. \ref{eq:pore} and the total epithelial permeability as a sum of the permeabilities of the dynamic strand breaking pathway and the tricellular tubes.

Tervonen \textit{et al.} \cite{tervonen2019} also simulated the electrical barrier using the same geometrical idea as in their permeability model, but calculated the total dynamic strand system resistance by solving the circuit formed into the strand network. For each current loop formed into the strand network, an equation of the following type was formed

\begin{equation}
\sum_j^n R_{ij}(t)I_i - \sum_{j \neq j}^n R_{ij}(t)I_j =
\begin{cases}
V_s, \quad &\text{if outer current loop}\\
0, \quad &\text{otherwise}
\end{cases},
\end{equation}
where $R_{ij}$ is a stochastic resistor shared by current loops $i$ and $j$, $I_i$ is the current in loop $i$, and $V_s$ is the measurement voltage. An average dynamic strand network resistance is calculated by solving the linear system of equations over time. Finally, the total TER can be calculated by connecting the resistance of the static tricellular pore in parallel with the resistance of the strand-breaking dynamics.

The resistive properties of the dynamic random strand network were studied by Washiyama \textit{et al.} \cite{washiyama2013} using percolation theory where the strand network was described by a random resistor network.

\subsubsection{Epithelial level (scale 3)}

In the tissue scale  (Fig.~\ref{fig:rule_ep}), the epithelial barrier is usually modeled as a homogeneous layer consisting of different pathways for electric current or molecules through it. The electrical properties of homogeneous epithelial tissue can be represented by equivalent circuits, where the components of the epithelial barrier are described using a combination of resistors and capacitors \cite{gunzel2012,gerasimenko2020}. In the simplest equivalent circuit, these properties are separated simply by connecting the resistive and capacitive properties in series and can thus be described by 

\begin{equation}
Z_{epi} = \frac{R_{epi}}{1+ i \omega C_{epi} R_{epi}},
\end{equation}
where $i=\sqrt{-1}$, $\omega$ is the angular frequency, and $R_{epi}$ and $C_{epi}$ are the resistance and capacitance of the epithelium, respectively \cite{savolainen2011,onnela2012}. Also, more complex circuits have been used, for example, those that further divide the resistance into paracellular and transcellular resistances \cite{gunzel2012}.

Similar strategies have also been used to study steady-state epithelial permeability. Here, the different barrier components can be connected in parallel or in series to describe the roles of the components in the epithelial barrier. For example, Edwards \& Prausnitz \cite{edwards2001} modeled the permeability of corneal epithelium for topical drugs by dividing the permeability into paracellular and transcellular pathways. A similar approach was also taken by Tervonen \textit{et al.} \cite{tervonen2014} to study the permeability of retinal pigment epithelium. These models usually take a form similar to the following example that considers the permeation through the paracellular pathway and a single transcellular pathway:

\begin{equation}
P_{epi} = P_{para} + P_{trans} = \left( \frac{1}{P_{tj}} + \frac{1}{P_{ls}} \right)^{-1} + \left( \frac{2}{P_{pm}} + \frac{1}{P_{cyt}} \right)^{-1}
\end{equation}
In this example, the paracellular permeability ($P_{para}$) consists of tight junctions ($P_{tj}$) and the space between the cells, or lateral space, ($P_{ls}$) permeabilities. Furthermore, the transcellular pathway is described by the permeabilities of the plasma membrane ($P_{pm}$) and the cytoplasm ($P_{cyt}$). The specific permeabilities of these components can be calculated based on usually phenomenological models that describe their permeability as a function of properties of the molecule, such as size and lipophilicity. For example, $P_{tj}$ can be described as small circular pores using Eq.~\ref{eq:pore}.

Similar approaches have also been taken to model the permeability of the epithelia in lungs \cite{yu2010,eriksson2018} and the skin \cite{mitragotri2003,schwobel2019,naegel2011,mitragotri2011}. For skin, the selected permeation pathways are generally different due to the distinct nature of the skin epithelium. Furthermore, time-dependent multicompartment and finite element models have been used for the permeation through the skin \cite{barbero2006,barbero2017,naegel2011,mitragotri2011}.

Another class of models describing the permeability of epithelial tissues is the \textit{quantitative structure-property relationship} models, where the permeability of the epithelium is defined phenomenologically based only on the properties of the diffusing molecules \cite{subramanian2006,cronin2007,kidron2010,pecoraro2019}. This technique is especially useful for quickly estimating the permeability properties of new drug molecules.

\subsubsection{Organ-level epithelial models (scale 4)}

While epithelia usually form a part of an organ rather than being an organ themselves, they have been included as a component of the model in organ-scale drug delivery studies (Fig.~\ref{fig:rule_ep}). In these models, usually taking the form of a multicompartment or finite element model, the properties of the epithelium are reduced to a single permeability coefficient. For example, in ocular drug delivery, the properties of the retinal pigment epithelium are usually described in this way \cite{macgabhann2007,amrite2008,ranta2010,balachandran2008,jooybar2014}.

\section{The future of \textit{in silico}  pharmacology}



Computational models have been used in drug development for many years, for drug discovery \cite{kuemmel2020consideration,thomas2009physiologically,kumar2006applying}, to safety pharmacology \cite{rusyn2010computational,davies2016recent,kugler2020modelling}, and toxicology \cite{kuemmel2020consideration,rusyn2010computational,kavlock2008computational}. Big pharmaceutical companies often have their own internal team of bioinformaticians who integrate computer science methodologies directly into their pipelines. Here, we aim to clarify why this partnership is far away from its full potential and highlight ongoing challenges for the widespread use of \textit{in silico} trials in pharmacology as an effective alternative to current methodologies. It is important to point out that in recent years a lot of work has been done to promote the use of modeling and simulation across academia, industry, and regulatory agencies, and to define a framework to assess model credibility \cite{musuamba2021scientific,kuemmel2020consideration}.
We will consider cardiac safety as a representative example of how \textit{in silico} methods can bring innovation compared to the currently used methodologies. As described above, multiscale computer models of the heart have been used for more than 60 years. There are numerous human-based models publicly available, and they have been validated and used over the years in a variety of different contexts. Therefore, they constitute a “mature technology”, which is ready to go beyond academia and have a concrete impact in the industry. Those models can be used as an alternative or in combination with current methodologies, as described in more detail below. 
Drug-induced cardiotoxicity, i.e., adverse effects on the heart function, caused by a drug still constitutes a big challenge in drug development \cite{ferri2013drug,kelleni2018drug}. Pre-clinical testing – mainly performed in animal models – are not always able to predict the effects later observed in humans \cite{valentin2022challenges}. Since 2013, the need of rechanneling the cardiac proarrhythmia safety paradigm has started to emerge, also thanks to the launch of the CiPA (Comprehensive \textit{in vitro} Proarrhythmia Assay) initiative, promoted by the FDA and other organizations \cite{sager2014rechanneling,gintant2016evolution}. CiPA includes \textit{in silico} cardiac cell models as one of its primary components, together with the use of human stem cell-derived cardiomyocytes. This led to a global effort that resulted in numerous publications, all aimed at demonstrating the power of computer modeling and simulations for drug cardiac safety, using different methodologies \cite{passini2017human,li2019assessment,llopis2020silico,varshneya2021prediction,fogli2021sex}. Most of these studies have been performed at the cellular level, to maximize performances in standard computers and provide fast predictions, even if there are some examples at the whole heart level \cite{sahli2018predicting}. Two user-friendly software are also available to perform cardiac safety simulations: i) the Virtual Assay (Oxford University Innovation\copyright~2018)\cite{passini2021virtual}, and the Cardiac Safety Simulator (CSS)  by Certara\textregistered.
Thanks to these validation studies that raised the profile of computer modeling and simulations for drug cardiac safety, more and more companies started to incorporate \textit{in silico} evidence in their publications \cite{morissette2020combining,delaunois2021applying,qu2021comprehensive}. This suggests that changes are possible, even if sometimes their implementation is somehow slow.

To evaluate the readiness of multiscale models in pharmacology, we have chosen three tissue types (heart, brain, and epithelium) and their respective available computational models. Table~\ref{tab:2} provides an overview of the current multiscale modeling strategies, the level of complexity and predictability as well as the gap to clinical application and the level of personalization.

However, there are still ongoing challenges to address for full integration of \textit{in silico} methods in pharmacology. Here, we decided to focus on two crucial ones. First, there is no clear standard for the collection of the input data used to characterize the effect of the drugs at the cellular level. These data are used to construct, calibrate, and validate \textit{in silico} models, and their consistency is really important to obtain accurate predictions. The Ion Channel Working Group (ICWG), developed within the CiPA initiative is working to deliver best practice recommendations for generating these data \cite{fermini2016new}. However, at the moment, there is a large variability between different laboratories and even within the same laboratories over time.
Second, there are many other sources of variability that influence drug response in humans, e.g., genetic profiles, sex, hormones, underlying conditions, and concomitant medications. Taking these factors into account is key to predicting drug effects at the population level.
Over the years, multiple strategies have been proposed to incorporate different sources of variability in computer modeling and simulations \cite{fogli2021sex,passini2017human,chang2017uncertainty}, even if there is no regulatory recommendation of which would be more appropriate to use until today.

\section{Discussion}

\begin{table*}[]
\centering
\caption{Cross-tissue qualitative evaluation of complexity, clinical application, and personalization}
\begin{tabular}{|c|c|c|c|}
                            \hline                      & Cardiac & Brain & Epithelia \\ \hline  \hline  
Multiscale modelling strategy &  Mostly Biophysical & Biophysical and Phenomenological &    Solely Biophysical       \\ \hline  
Level of complexity                               &     high    &  high      &     high      \\ \hline
Level of predictability (risk)                             & high  &  low     & low          \\ \hline 
Gap to clinical application                       &     low    &   high     &  high  \\\hline 
Level of personalization                       &        low &   low     & low \\\hline 
\end{tabular}
\label{tab:2}
\end{table*}

Throughout scales and tissue types, the balance between prediction accuracy and modeling complexity interferes with the ability to describe biology. One needs to recognize that the computational infrastructure to have a full working version of a descriptive model of tissues or organs needs to be of a full-scale data center. For example, the \textit{Blue Brain Project} researchers work with data centers of IBM. They use highly sophisticated biological models to describe the workings of compartments of neurons (a 20$\mu$m neuron can have more than 50 compartments) along with their ion channels, in order to build realistic neuronal activity from a realistic morphological description of each neuron. However, the spatial scale only covers from the $\mu$m and above. Thus, the interactions of individual ions, proteins, and molecular structures are not fully accounted for. As the spatial scale grows, more computational resources are needed to retain the spatial and functional accuracy with increasing simulation runtime, which tends to make researchers consider less and less precise biological descriptions and focus on the "grey boxes" of phenomenological models. Phenomenological models consist of approximated or probabilistic approaches to account for groups of cells and their communications in a computationally feasible way. Those methods do, however, produce reliable results and the issue is not about their reliability in the scale they operate. When attempting to integrate different multiscale models, they tend to not provide a representative model that is also descriptive. 

The description of biological phenomena is largely reduced to particular events by which probabilistic volume in models is used. Models should always strive to provide realistic yet computationally feasible descriptions. We argue that a combination of biophysical and phenomenological models needs to be further explored in the future for multiscale modeling. Alternatively, the development of models must have an integration for other scales as a requirement, which allow seamless integration. Such an approach still needs to have a biophysical basis for smaller scales in building two phenomenological models at the largest scales. That has been a successful approach in heart modeling. Researchers now have the capability of inferring a whole organ state based on changes in ionic channels from scale levels in space and time. However, a reversible approach of having higher-level scale information be used to predict lower-scale information may seem a hard challenge if you consider the mixture of biophysical and phenomenological models.  

Since the whole idea of using digital twin solutions is to provide certain reliable levels of prediction based on the response of adding a drug into a tissue or model, we argue that this predictive ability should be maintained in a computationally feasible solution so that it can become a service. We hope that initiatives for distributed computing or high-performance computing can be applied to integrate the complexity of the lower scale levels with accurate phenomenological models of higher scale levels. Computer scientists have been amazingly successful at administrating large quantities of data and models in the context of data centers, and know how to exploit this infrastructure correctly. However, in the integration of biophysical and phenomenological models, there may be opportunities to apply old techniques of high-performance computing or develop new ones dedicated to the type of modeling and data that we see in biophysical models.

Advancements in the area of biophysics take time since they have to deal with the full validation of drug modeling integration to the existing models. Novel methods are likely needed to deal with drug development computationally in an accessible framework facilitating testing in a digital twin solution with full confidence in representing the biological desired prediction. Such models will need to take into account different levels of the physical environment. In that way, biophysical modeling needs to address pending challenges such as the variations of pressure, temperature, or other changing factors that interfere with the most basic bricks at a lower scale level. That is, however, not to exclude the existing reliable models. In lieu, to employ \textit{in silico} methods in the drug validation pipeline, the constant effort for improving prediction accuracy should be driven by, but not limited to, the ion channels and any other spatial temporal scale below that.

Several initiatives have been established to store models and to standardize data. In general, platforms like CellML \cite{cellml} allow sharing of computational models of various cell types. The website of COMBINE gives minimum information about conducting simulation experiments \cite{combine1} and minimal information required for model annotation \cite{combine2}. In computational neuroscience, for example, ModelDB \cite{ModelDB} and INCF \cite{INCF} provide a platform for sharing models as well as data reuse and reproducibility, respectively.

\section{Conclusion}

Multiscale biophysics needs to be addressed before we implement digital twin solutions. We argue that overall, across all tissue types, there is a long way to go before the modeling of multiscale biophysics is considered mature enough to draw precise prediction measures. Cardiac models have a level of maturity that is perhaps closest to achieving prediction, as evidenced by the recent development at the research community level towards the usage of their models in pharmacological \textit{in silico} solutions with already significant industry involvement. In our review, we not only expose biophysical models from the cardiac, brain, and epithelial tissue types but highlight their drawbacks through the lenses of multiscale modeling and argue that many lessons from cardiac modeling can be incorporated into the modeling of other tissue types to advance refereed models to a similar level of maturity. The main goal should be reliable and predictive modeling that can become a service for industry and academia to optimize the drug development pipeline and avoid the future need for animal trials. Both points may lead to a dramatic decrease in route-to-marked costs of drugs, which could be centered on powerful multiscale modeling.

\bibliographystyle{unsrt}  
\bibliography{references}

\end{document}